\documentclass[reqno,12pt]{article}
\pdfoutput=1
\usepackage{ae}
\usepackage[T1]{fontenc}
\usepackage[ansinew]{inputenc}
\usepackage{mathrsfs}
\usepackage{amsmath}
\usepackage{dsfont}
\usepackage{amssymb}
\usepackage{graphicx}
\usepackage{color}
\definecolor{darkblue}{cmyk}{0.9,0.9,0,0}
\definecolor{darkgreen}{rgb}{0,0.55,0}
\usepackage[colorlinks=true,linkcolor=darkblue,citecolor=darkblue,urlcolor=darkblue]{hyperref}
\usepackage{epsfig}
\usepackage{cite}
\usepackage{mathtools,slashed}
\usepackage{caption}
\usepackage{subcaption}
\usepackage{cancel}

\usepackage{amsmath,amsfonts,amssymb,amsbsy}
\usepackage{graphicx,xcolor}
\usepackage{graphics}
\usepackage{epsfig}
\usepackage{verbatim}
\usepackage{stmaryrd}
\usepackage{epsf}

\usepackage{amsmath, amssymb,graphicx}
\usepackage{epsfig}
\usepackage{verbatim}
\usepackage{amsfonts}
\usepackage{ulem}
\usepackage{wrapfig}
\usepackage{tikz}
\numberwithin{equation}{section}

\usepackage{tikz}
\usetikzlibrary{arrows,shapes}
\usetikzlibrary{patterns,fadings}
\usetikzlibrary{decorations.pathmorphing}

\newcommand{\btp}{\begin{tikzpicture}[baseline=0pt,scale=0.9,line width=0.7pt]}
\newcommand{\btpp}{\begin{tikzpicture}[baseline=-5pt,scale=0.25,line width=0.7pt]}
\newcommand{\etp}{\end{tikzpicture}}

\def\bc{\begin{center}}
\def\ec{\end{center}}

\usepackage{epsf}

\newcommand{\be}{\begin{equation}}
\newcommand{\ee}{\end{equation}}
\newcommand{\ba}{\begin{eqnarray}}
\newcommand{\ea}{\end{eqnarray}}

\newcommand{\beaa}{\begin{eqnarray}}
\newcommand{\eeaa}{\end{eqnarray}}

\newcommand{\ket}[1]{{\,\left|#1\right>}\,}

\definecolor{darkgreen}{rgb}{0.1,0.7,0.1}

\DeclareFontFamily{OT1}{pzc}{}
\DeclareFontShape{OT1}{pzc}{m}{it}{<-> s * [1.10] pzcmi7t}{}
\DeclareMathAlphabet{\mathpzc}{OT1}{pzc}{m}{it}

\def\({\left(}
\def\){\right)}
\def\[{\left[}
\def\]{\right]}

\def\<{\langle}
\def\>{\rangle}

\def\s*{\ *_{\!\!\!\!\!\!\!\!\!\,_{\,_\text{\scriptsize{sym}}}}}
\def\hs*{\ \hat{*}_{\!\!\!\!\!\!\!\!\!\,_{\,_\text{\scriptsize{sym}}}}}

\begin{document}

\thispagestyle{empty}

\renewcommand{\thefootnote}{\fnsymbol{footnote}}
\setcounter{footnote}{0}
\setcounter{figure}{0}

\begin{center}
$$$$

\vspace{1.0cm}

{\Large\textbf{\mathversion{bold}
Boundary bound states and \\
 integrable Wilson loops in ABJM
}\par}

\vspace{1.0cm}

\vspace{1.0cm}

\textrm{Diego H. Correa\footnote{correa@fisica.unlp.edu.ar}, Maximiliano G. Ferro\footnote{mgferro.vl@gmail.com}, Victor I. Giraldo-Rivera\footnote{vigirald@gmail.com} and \\ 
Nicol\'as A. Ivanovich \footnote{nicolas.ivanovich@fisica.unlp.edu.ar}}
\\ \vspace{1.2cm}
\footnotesize{

\textit{Instituto de F\'{\i}sica La Plata, Universidad Nacional de La Plata, \\ C.C. 67, 1900 La Plata, Argentina} \\
\texttt{} \\
}

\par\vspace{1.5cm}

\textbf{Abstract}\vspace{2mm}
\end{center}

\noindent

We derive an integrable reflection matrix  for the scattering of excitations from a boundary with a degree of freedom when the reflection process preserves an $SU(1|2)$ symmetry. As this residual symmetry is not sufficient to fully determine the reflection matrix, we use the boundary remnant of the Yangian symmetry invariance and obtain a family of integrable solutions. A concrete realization of this setup is found when studying insertions in the 1/2 BPS Wilson loop in ABJM theory. The boundary degree of freedom appears as a boundary bound state due to poles in the dressing phase of the reflection matrix. We also compare our results with those obtained from the boundary bound state bootstrap procedure. The ABJM Wilson loop example enables us to perform perturbative verifications of our results.

\vspace*{\fill}

\setcounter{page}{1}
\renewcommand{\thefootnote}{\arabic{footnote}}
\setcounter{footnote}{0}

\newpage

\section{Introduction}
\label{introduction}

Wilson loops in superconformal gauge theories constitute foremost examples of line defects, as they encode important physical information about the theories themselves.  For suitable choices, Wilson loop operators define one-dimensional defect superconformal field theories, and correlation functions of excitations inserted along these line defects are basic observables to study \cite{Drukker:2006xg}.

These correlation functions  can be studied using a variety of mathematical tools. In the well-established cases of $\mathcal N=4$ super Yang-Mills and $\mathcal N=6$ super Chern-Simons-matter theories, integrability provides a powerful approach to systematically study the spectrum of scaling dimensions of excitations on conformal line defects.

In ${\cal N}=4$ super Yang-Mills, the connection between the dilatation operator on the Wilson lines with insertions  and the Hamiltonian of certain integrable open spin chains was first noted in a perturbative computation \cite{Drukker:2006xg}. Later on, the all-loop reflection matrix was determined using symmetry arguments and a Boundary Thermodynamic Bethe Ansatz was derived for the exact computation of scaling dimensions in the planar limit \cite{Drukker:2012de,Correa:2012hh}. It is important to emphasize that integrability is not an assumption, but rather one discovers that the bootstrapped reflection matrix satisfies the boundary Yang-Baxter equation (BYBE). Among many relevant related developments, we can mention the Quantum Spectral Curve formalism, which can be combined with conformal bootstrap ideas to efficiently access finite coupling regimes \cite{Gromov:2013pga,Gromov:2015dfa,Liendo:2018ukf,Cavaglia:2021bnz}.

The extension of these ideas to study Wilson loops in the ${\cal N}=6$ super Chern-Simons-matter theory was initiated in \cite{Correa:2023lsm}. In this case,  the reflection matrix off the boundaries imposed by the 1/2 BPS Wilson line is constrained by an $SU(1|2)$ symmetry. This symmetry group is the common subgroup of the reference state's $SU(2|2)$ and the Wilson loop's $SU(1,1|3)$. For magnon excitations in the fundamental representation of $SU(2|2)$, the underlying residual symmetry is enough to fix the reflection matrix, which also satisfies the BYBE. This is a clear indication of the integrability of the spectral problem for insertions on the 1/2 BPS Wilson line. This is, in many regards, similar to the analogous problem in the  ${\cal N}=4$ super Yang-Mills theory. However, given that $SU(1|2)$ is a relatively small group, its symmetry constraints may be insufficient to determine the reflection matrix completely for excitations in other representations. While for $Q$-magnon bound states in totally-symmetric representations the residual $SU(1|2)$ symmetry is enough to determine the reflection matrix,
for $Q$-magnon bound states in totally-antisymmetric representations it is not. Given the evidence of integrability in the fundamental representation, one can complete the determination of these other reflection matrices, by demanding to be those of an integrable boundary condition.

In that context, a powerful tool can be the use of Yangian symmetry, an infinite-dimensional extension of the underlying Lie (super)algebra symmetry that governs the integrable structure. Indeed, as recognized early in \cite{Beisert:2006fmy}, the $su(2|2)$ symmetry algebra from which the bulk S-matrix is derived admits a Yangian extension  $Y({su}(2|2))$. This extension has become a central tool  for the construction of the S-matrix for general higher-rank representations \cite{Arutyunov:2009mi}, where ordinary symmetries alone are insufficient (see \cite{Loebbert:2016cdm} for a review and \cite{Loebbert:2024qbw,Loebbert:2025abz} for recent connections to scattering amplitudes). The Yangian extension can also be performed for the residual symmetry that leaves both a boundary and the reference state invariant \cite{MacKay:2001bh,Delius:2001he}. This turned out to be  crucial in determining reflection matrices for magnon bound states in the $Y=0$ maximal giant graviton problem \cite{Hofman:2007xp,Ahn:2010xa,MacKay:2010zb,MacKay:2010ey,Palla:2011eu}. 

In the present work, we apply these ideas to characterize a new feature that appears in the Wilson loop reflection matrix of ${\cal N}=6$ super Chern-Simons-matter theory: a pole in the overall dressing factor of the reflection matrix, which suggests the existence of boundary bound states excitations trapped at the boundary \cite{Correa:2023lsm}.

Our goal is to determine the reflection matrix when a magnon excitation is already trapped at the boundary. Although our original motivation stems from the Wilson loop setup just described, we will consider reflection matrices which could describe more general systems. As we shall see, the residual $SU(1|2)$  symmetry is not enough to fix this reflection matrix completely. As the original problem was integrable, it seems natural to impose the BYBE to further constrain the reflection matrix. Alternatively, as this procedure would become rather involved, we will require that a boundary remnant of the bulk Yangian symmetry is preserved. This condition provides additional constraints that fully determine the reflection matrix, leveraging the integrable structure underlying the problem.

The paper is organized as follows. In Section \ref{IntegrableWL}, we provide a brief review of the integrable open spin chains used to study the spectral problem for insertions in the 1/2 BPS Wilson loop of the ABJM model and we review how to use symmetries to constrain the scattering of excitations propagating on a suitable reference state. In Section \ref{R-matrix vac}, we consider the case of a boundary invariant under the residual $SU(1|2)$ symmetry. After briefly reviewing the reflection of excitations in the fundamental representation, we generalize the analysis by considering the reflection of bound states of $Q$ magnons. The poles in the corresponding dressing factor reveal the existence of boundary bound states in higher-rank representations. Then, in Section \ref{R-matrix with d.o.f}, we focus on the reflection of excitations off a boundary that contains a degree of freedom and observe that the residual $SU(1|2)$ symmetry is not enough to fix it completely. In Section \ref{boundary yangian}, we assume the existence of an infinite set of conserved charges associated with the Yangian extension of the $su(1|2)$ symmetry preserved by the boundary and use it to complete the determination of the reflection matrix for the case with a boundary degree of freedom. Finally we discuss how our reflection matrix derivation is related to the boundary bound state bootstrap procedure. In Section \ref{weak} we perform several calculations with the perturbative open spin chain Hamiltonian to verify the consistency of the weak-coupling limit of our all-loop results. Finally, we discuss our findings and future directions in Section \ref{discu}.

\section{$SU(1|2)$ integrable boundaries}
\label{IntegrableWL}
 
Examples of bulk spin chains, whose $SU(2|2)$ symmetry is reduced to $SU(1|2)$ by the presence of a boundary have been found in various AdS/CFT setups, both in ${\cal N}=4$ super Yang-Mills and ${\cal N}=6$ super Chern-Simons-matter theories 
\cite{Hofman:2007xp,Correa:2008av,Chen:2018sbp,Correa:2023lsm}. In all those examples, the boundary is in a trivial representation of the residual $SU(1|2)$ symmetry.

This article initiates the study of boundaries with degrees of freedom transforming under a non-trivial representation of $SU(1|2)$. Having a concrete example, as the one we introduce in what follows, offers a valuable first approximation to this problem, as it would allow us to verify some of our results by comparison with explicit computations.

Let us review the 1/2 BPS Wilson line of the ${\cal N}=6$ super Chern-Simons-matter theory \cite{Aharony:2008ug} (ABJM model from now on), discuss its symmetries and how the scaling dimension of local operator insertions can be described through an integrable open spin chain \cite{Correa:2023lsm}. 
Wilson loops in ABJM can be defined in terms of the holonomy of a $U(N|N)$ superconnection \cite{Drukker:2009hy,Lee:2010hk}
\begin{equation}
\label{WLdef}
 W=
 \frac{1}{2N} \operatorname{Tr}\left[\mathcal{P} \exp \left(i \int
 \mathcal{L}(\tau) d \tau\right)\right]\,,
\end{equation}
with 
\begin{equation}
    \mathcal{L}=\left(\begin{array}{cc}
A_\mu \dot{x}^\mu-i|\dot{x}| M_J^I C_I \bar{C}^J & -i|\dot{x}| \eta_I^\alpha \bar{\psi}_\alpha^I \\
-i|\dot{x}| \psi_I^\alpha \bar{\eta}_\alpha^I & \hat{A}_\mu \dot{x}^\mu-i|\dot{x}| M_J^I \bar{C}^J C_I
\end{array}\right),
\label{Ldef}
\end{equation}
and where $M^{I}_{J}$ and $\eta^{\alpha}_{I}$ are Grassmann-even couplings. For a straight line parametrized by $x^\mu=(\tau, 0,0)$ and Grassmann-even couplings given by
\begin{equation}
    M_J^I=-\delta_J^I+2 \delta_1^I \delta_J^1, \quad \eta_I^\alpha=\eta \delta_I^1 \delta_{+}^\alpha, \quad \bar{\eta}_\alpha^I=\bar{\eta} \delta_1^I \delta_\alpha^{+}\,,
    \label{orientations}
\end{equation}
the Wilson loop \eqref{WLdef} turns out to be 1/2 BPS  provided $ \eta\bar{\eta}=-2i$. We have moved to the appendices \ref{app: ABJM conventions} and \ref{app: review WL} the conventions of the ABJM gauge theory and a detailed analysis of the supersymmetric properties of the Wilson loops and their local operator insertions. The supersymmetries, altogether with the bosonic symmetries preserved by the line, define an $SU(1,1|3)$ supergroup, a remnant of the $OSp(6|4)$ symmetry of the ABJM model. 

Generic local operator insertions along the 1/2 BPS Wilson line constitute the excitations in the conformal line. Correlation functions among them are simply obtained by computing expectation values of the Wilson line
with the local operator insertions. For instance, the 2-point correlation function is
\begin{equation}
 \langle\!\langle {\cal O}^\dagger(\tau_1)
 {\cal O}(\tau_2) \rangle\!\rangle 
 :=  \frac{1}{2N} \Big\langle\operatorname{Tr}\Big[\mathcal{P} \exp \left(i \int
 \mathcal{L}(\tau) d \tau\right) {\cal O}^\dagger(\tau_1)
 {\cal O}(\tau_2)\Big]\Big\rangle
 \propto \frac{1}{|\tau_2-\tau_1|^{2\Delta}}
 \,,
 \label{2ptintheline}
\end{equation}
and can be used to compute the scaling dimension of the insertion ${\cal O}$.

To analyze the spectrum of scaling dimensions, the identification of supersymmetric insertions turns out to be quite useful. Since they belong to short representations, their scaling dimensions are protected and do not receive quantum corrections. Thus, they can be regarded as reference states, in the sense that more general insertions can be accounted as excitations on them. An example of such supersymmetric insertion is  \cite{Correa:2023lsm}.
\be 
{\cal O}_\ell = 
\left(
\begin{array}{cc}
 0 &  C_1 (\bar C^2 C_1)^\ell
\\
0 &  0
\end{array}\right)\,,
\label{susyinsertion0}
\ee
which breaks the $SU(1,1|3)$ symmetry of the Wilson loop down to $SU(1|2)$. 

The off-diagonal block, of the form $C_1 \bar C^2\cdots C_1\bar C^2 C_1$, is a usual chain of ABJM scalar fields. If it were not inserted in the Wilson loop, it would be invariant under $SU(2|2)$ transformations. More general operators can be thought as impurities in the supersymmetric reference operator \eqref{susyinsertion0}. Depending on whether the impurities replace a $C_1$ or a $\bar C^2$, they are classified into types $A$ and $B$.

The action of the dilatation operator, computed from \eqref{2ptintheline}, mixes different operator insertions but with the same number and types of impurities and can be regarded as the Hamiltonian of an open spin chain \cite{Minahan:2008hf,Gaiotto:2008cg,Gromov:2008qe}. To diagonalize this Hamiltonian it is customary to consider magnons, {\it i.e.} waves of impurities  with certain momentum $p$. In this picture, the Wilson loop defect acts as a boundary for the spin chain, and the reflection of magnons from this boundary is encoded in a reflection matrix.

The diagonalization of this open boundaries Hamiltonian appears to be an integrable problem in the planar limit \cite{Correa:2023lsm}. Thus, the resolution of this spectral problem can be achieved by means of a Bethe Ansatz, whose implementation would require knowing the bulk scattering matrix $S(p,q)$ and the reflection matrix $R(p)$.

The symmetry of the reference state is $SU(2|2)$ and turns out to be quite restrictive on the S-matrix of magnon excitations. Magnon impurities accommodate in representations of the centrally extended  $SU(2|2)$ group, which are characterized by four coefficients $(\mathsf a,\mathsf b,\mathsf c,\mathsf d)$.

For a magnon in the fundamental representation, the action of $SU(2|2)$ fermionic generators is given by 
\begin{alignat}{6}
{\mathbb Q}_{\alpha}^{\ a} \ket{\phi_b} & = {\sf a} \; \delta^b_a \ket{\psi_{\alpha}}\,, 
& & \qquad & 
{\mathbb Q}_{\alpha}^{\ a}\ket{\psi_{\beta}} & = {\sf b}\; \epsilon_{\alpha\beta}\epsilon^{ab}\ket{\phi_b}\,,
 \label{actionofQ}
\\
{\mathbb S}_{a }^{\ \alpha}\ket{\phi_b} &=  
{\sf c} \;\epsilon_{ab}\epsilon^{\alpha\beta}\ket{\psi_{\beta}}\,, 
 & & & 
{\mathbb S}_{a}^{\ \alpha}\ket{\psi_{\beta}}  &= {\sf d}\; \delta^{\alpha}_{\beta}\ket{\phi_a}
 \,,
 \label{actionofS}
\end{alignat}
where representation coefficients
are parametrized as 
\begin{equation}
    \mathsf {a}=\sqrt{g}\,\eta\,,\;\;\;\mathsf{b}=\sqrt{g}\,\frac{i\zeta}{\eta}\left(\frac{x^+}{x^-}-1\right)\,,\;\;\;\mathsf {c}=-\sqrt{g}\,\frac{\eta}{\zeta x^+}\,,\;\;\;\mathsf{d} =-\sqrt{g}\,\frac{x^+}{i\eta}\left(\frac{x^-}{x^+}-1\right)\,,
\end{equation}
with
\begin{equation}
    x^++\frac{1}{x^+}-x^--\frac{1}{x^-}=\frac{i}{g},\;\;\;\;\;\;\frac{x^+}{x^-}=e^{ip},\;\;\;\;\;\; \eta(\zeta,x^+,x^-)=\zeta^{1/2}e^{i\frac{p}{4}}\sqrt{i(x^--x^+)}\,,
\end{equation}
where $p$ is the magnon momentum and $\zeta$ is a phase. The coupling $g=g(\lambda)$ is a non-trivial interpolating function of the ABJM 't Hooft coupling. In the weak coupling limit
$g(\lambda) = \lambda -\frac{\pi^3}{3}\lambda^3 + {\cal O}(\lambda^5)$ \cite{Gromov:2014eha}.

One of the central extensions of the $SU(2|2)$ algebra directly gives the scaling dimension and it is related to the momentum through a dispersion relation 
\begin{equation}
    \mathbb{H} = E(p) 
   = \frac{g}{2i}\left(x^+ - \frac{1}{x^+} - x^- +\frac{1}{x^-}\right) 
    =\frac{1}{2}\sqrt{1+16g^2\sin^2(p/2)}.
    \label{dispersion}
\end{equation}

The S-matrix acts on the direct product of two such fundamental representations. By demanding it commutes with all $SU(2|2)$ generators, the S-matrix can be fixed up to an overall dressing phase \cite{Beisert:2005tm,Minahan:2008hf,Gaiotto:2008cg}. The scattering matrices between magnons of the same or different types differ only by a dressing phase.
\begin{equation}
    \begin{aligned}
        S^{AA}(x_1^\pm,x_2^\pm)&=S^{BB}(x_1^\pm,x_2^\pm)=S_0(x_1^\pm,x_2^\pm)\hat S(x_1^\pm,x_2^\pm),\\
        S^{AB}(x_1^\pm,x_2^\pm)&=S^{BA}(x_1^\pm,x_2^\pm)=\tilde S_0(x_1^\pm,x_2^\pm)\hat{S}(x_1^\pm,x_2^\pm),
    \end{aligned}
\end{equation}
where $\hat S(x_1^\pm,x_2^\pm)$ is the $SU(2|2)-$invariant matrix and $S_0$ and $\tilde S_0$ are the corresponding dressing phases fixed by  crossing symmetry \cite{Beisert:2006ez,Ahn:2008aa}. The conventions we use are spelled out in appendix
\ref{convencionessmatrix}.

\subsection{Reflection from a singlet boundary} 
\label{R-matrix vac}

As stated before, we are primarily interested in the case of boundary degrees of freedom in non-trivial representations. However, it is instructive to begin our analysis with the case of a singlet boundary. It is the presence of poles and zeros in the corresponding reflection matrix that gives rise to boundary bound states.

In the case of a singlet boundary, the residual symmetry $SU(1|2)$ is enough to determine the right reflection matrix for the two types of magnon excitations up to overall dressing factors \cite{Drukker:2019bev},
\begin{equation}
\begin{aligned}
R^{A} = &\  R^{A}_0 \, {\rm diag}(1,1,e^{-ip/2},-e^{ip/2}) \,,
\\
R^{B}  = &\  R^{B}_0 \, {\rm diag}(1,1,e^{-ip/2},-e^{ip/2}) \,.
\end{aligned}
\end{equation}
Imposing crossing symmetry and consistency with explicit weak and strong coupling computations leads to a proposal for the exact form of the dressing factors \cite{Correa:2023lsm},
\begin{equation}
\label{crossing ansatz}
\begin{aligned}
R^{A}_0(p) &= - {R_0 (p)} \left( \frac{\frac{1}{x^-}+x^-}{\frac{1}{x^+}+x^+} \right) \left( \frac{x^+}{x^-} \right)\,, \\
R^{B}_0(p) &= {R_0 (p)} \, \left( \frac{x^+}{x^-} \right)\,,  \\
\end{aligned}
\end{equation}
where $R_0(p)$ is the square root of the dressing factor proposed for the ${\cal N}=4$ super Yang-Mills Wilson line reflection matrix 
\cite{Correa:2012hh,Drukker:2012de}. Because of the extra factor of $\frac{\frac{1}{x^-}+x^-}{\frac{1}{x^+}+x^+}$ in $R^{A}_0(p)$, the reflection matrix of type $A$ magnons has a pole and a zero, which implies that they can form boundary bound states \footnote{In our conventions, the dissociation (association) amplitude of a boundary bound state is identified with a pole (zero) of the reflection matrix.} \cite{Correa:2023lsm}.  

To become trapped at the right boundary, the magnon must have spectral parameters
\begin{equation}
  x^- = -i\,, \qquad x^+ = x_B\,
  \qquad\text{where}\quad
  x_B = \frac{i}{2g}
  \left(1+\sqrt{1+4g^2}\right)\,.
\label{expaQ1bbs}  
\end{equation}
According to \eqref{dispersion}, the scaling dimension of such boundary bound state is
\begin{align} \label{exactenergypole}
E_B = &\  g +\frac12 \sqrt{1+4g^2}
\nonumber
\\
= &\ \frac{1}{2} + \lambda +\lambda^2 +{\cal O}(\lambda^3)\,,
\end{align}
where we have included its weak coupling expansion for later reference. The 1/2 is the classical scaling dimension and the remainder is the anomalous dimension\footnote{More precisely, the energy of the spin chain state is related to the difference of the scaling dimension and the R-charge of the reference state.}.

A possible generalization of the previous result is to consider the reflection of bound states of $Q$ magnons, transforming in totally symmetric $4Q$-dimensional representations of $SU(2|2)$. They also come in two types, depending on whether their constituents are type $A$ or type $B$
\cite{Bombardelli:2009xz}. If we label these multiplets components by $Q$ fundamental indices, the first $Q+1$ components would be $|\phi_{\{a_1\cdots a_Q\}}\rangle$ followed by $Q-1$ components $|\phi_{\{a_1\cdots a_{Q-2}\}[\alpha\beta]}\rangle$. Then, the last $2Q$ components would be fermionic $|\psi_{\{a_1\cdots a_{Q-1}\}\alpha}\rangle$. Demanding that its action commutes with the generators of the residual $SU(1|2)$ symmetry, the resulting reflection matrix is also diagonal and fixed up to an overall scalar factor\footnote{This turns out to be analogous to the reflection of bound states transforming in totally antisymmetric representations, in the case of a residual $SU(2|1)$ symmetry \cite{Correa:2009mz}.}
\begin{equation}
R_Q^A =  R_{Q0}^A\ {\rm diag}(\overbrace{\vphantom{e^{\frac{i}{2}p}}1,\cdots,1}^{Q+1},
\overbrace{-1\vphantom{e^{\frac{i}{2}p}},\cdots,-1}^{Q-1},
\overbrace{e^{-\frac{i}{2}p},\cdots,e^{-\frac{i}{2}p}}^{Q},
\overbrace{-e^{\frac{i}{2}p},\cdots,-e^{\frac{i}{2}p}}^{Q})\,,
\end{equation}
and similarly for $R_Q^B$.

The overall scalar factors can be related to those in \eqref{crossing ansatz} with a fusion principle, if we think of the bound state constituted by
$Q$ elementary
magnons with spectral parameters $(x_1^+,x_1^-), \dots, (x_Q^+,x_Q^-)$
such that
\begin{equation}
x_1^- = x^-\,, \quad x_1^+ = x_2^-\,, \quad \cdots\quad  x_Q^+ = x^+\,.    
\end{equation}
where now $x^\pm$ satisfies the $Q$-magnon constraint
\begin{equation}
x^{+}+\frac{1}{x^{+}}-x^{-}-\frac{1}{x^{-}}=\frac{iQ}{g}\,.
\label{14:mass-shell_Q}
\end{equation}
The scalar factor for a totally symmetric representation reflection
matrix is 
\begin{equation}
R_{Q0}^{A} = \prod_{i=1}^{Q} R^{A}_0(x_i^\pm) \prod_{1\leq j<k\leq Q} S_0(x_j^\pm,-x_k^\mp)\,,
\label{fusionRQ}
\end{equation}
where $R_0^A$ and $S_0$ are the dressing factors given in \eqref{crossing ansatz} and 
\eqref{S0AA}.
It is worth noting that the first product in \eqref{fusionRQ} becomes simply
\begin{equation}\label{Qreflex0}
\prod_{i=1}^{Q} R^{A}_0(x_i^\pm) = (-1)^Q R^{A}_0(x^\pm) =    
(-1)^Q{R_0 (x^\pm)} \left( \frac{\frac{1}{x^-}+x^-}{\frac{1}{x^+}+x^+} \right) \left( \frac{x^+}{x^-} \right)\,,
\end{equation}
which still presents a pole and a zero. In this case, to become trapped at a right boundary, the magnon must carry the following kinematic variables:
\begin{equation}
  x^- = -i\,, \qquad x^+ = x_B^{(Q)}\,,
  \qquad\text{where}\quad
  x_B^{(Q)} = \frac{i}{2g}\left(
  Q+\sqrt{Q^2+4g^2}\right)\,.
\label{expaQQbbs}  
\end{equation}

The scaling dimension of these 
higher-rank representation boundary bound states is
\begin{align}\label{exactenergypoleQ}
E_B^{(Q)} = &\  g +\frac{1}2 \sqrt{Q^2+4g^2}
\nonumber
\\
= &\ \frac{Q}{2} + \lambda +\frac{\lambda^2}{Q} +{\cal O}(\lambda^3)\,,
\end{align}
which is valid for any integer $Q\geq 1$.

\section{Reflection from a boundary with a degree of freedom} 
\label{R-matrix with d.o.f}

As we have seen, even if one starts with a boundary that is a singlet of the residual symmetry, the presence of poles and zeros in the reflection matrix indicates the possibility of exciting the boundary with magnon degrees of freedom. It is therefore natural to enquire about the reflection of propagating magnons from a boundary that has a trapped degree of freedom. 
These poles are present in the reflection matrix of magnons in different representations, then boundary degrees of freedom can exist for representations of $SU(1|2)$ of different sorts.  For concreteness, we shall focus on a 4-dimensional representation that originates as a boundary bound state formed from a magnon in the fundamental representation of the bulk $SU(2|2)$ symmetry.
The action of the fermionic generators ${\mathbb Q}_{3}^{\ a}$ and  ${\mathbb S}_{a}^{\ 3}$ on the components of the state at the boundary would still be of the form given in \eqref{actionofQ} and \eqref{actionofS}, but with coefficients  
\begin{equation}
\mathsf{a}_B=\sqrt{g}\,\sqrt{\zeta}\eta_1\,,\;\;\;\mathsf{b}_B=\frac{\sqrt{g}\,\sqrt{\zeta}}{\eta_2}\,,\;\;\;\mathsf{c}_B=\frac{i\sqrt{g}\,\eta_3}{\sqrt{\zeta}}\,,\;\;\;\mathsf{d}_B=-\frac{i\sqrt{g}}{\sqrt{\zeta}\eta_4}\,,
\end{equation}
where $\zeta$ is a phase and the constraint
\begin{equation}
  \mathsf{a}_B \mathsf{d}_B-\mathsf{b}_B\mathsf{c}_B =1  
    \quad \Leftrightarrow\quad
    \eta_4=\frac{g\eta_1\eta_2}{i\eta_2-g\eta_3}\,,
\end{equation}
is needed to ensure the closure of the algebra. The energy of the boundary magnon is
\begin{equation}
    E_B = \frac12 (\mathsf{a}_B \mathsf{d}_B+\mathsf{b}_B\mathsf{c}_B)\,.
\end{equation}
Although we could further restrict the representation parameters to match \eqref{exactenergypole},  we will keep the value of the energy arbitrary at this point.

For the case of a fundamental boundary magnon, the reflection matrix  operates on the direct product of two fundamental representations\footnote{In (\ref{Rdof}), magnons are in fundamental representations of the original $su(2|2)$. Let us recall that we are using $(\phi^a|\psi^\alpha)=(\phi^1,\phi^2,\psi^3,\psi^4)$ for the fundamental of $su(2|2)$. These $(2|2)$-dimensional representations can be seen as graded 2-symmetric representations of $su(1|2)$. Given a $(\chi|\varphi^a)$ fundamental representation of $su(1|2)$, the components of a graded 2-symmetric representation would be $(\chi\chi,\varphi^1\varphi^2-\varphi^2\varphi^1,\varphi^1\chi-\chi\varphi^1,\varphi^2\chi-\chi\varphi^2)$ (See \cite{Hofman:2007xp} for a similar discussion).}: the propagating and the trapped magnons. 
\begin{equation}
      R : \ \ \boxslash_{(\zeta,p)} \otimes \boxslash_{(\zeta e^{ip})}\ \rightarrow \ \boxslash_{(\zeta,-p)}\otimes \boxslash_{(\zeta e^{-ip})}\,,
\label{Rdof}
\end{equation}
where we have indicated how the right reflection changes both the momentum and the phases of the magnons.

In order to preserve the residual symmetry, the reflection matrix has to commute with the $su(1|2)$ generators.
The most general reflection matrix commuting with the bosonic generators can be expressed in terms of twenty functions,
\begin{align} 
&  {R}_{ab}^{cd} = \tfrac{1}{2}(A_B+B_B) \delta_a^c\delta_b^d + \tfrac{1}{2}(A_B-B_B)   \delta_a^d\delta_b^c\,,
\nonumber
  \\
&{R}_{ab}^{\alpha\beta} =  -\tfrac{1}{2}\epsilon_{ab}
\big(\delta_3^\alpha\delta_4^\beta C_B^1 - \delta_4^\alpha\delta_3^\beta C_B^2\big)\,,
\nonumber
  \\
&{R}^{ab}_{\alpha\beta} =  \tfrac{1}{2}\epsilon^{ab}
\left(\delta^3_\alpha\delta^4_\beta F_B^1 - \delta^4_\alpha\delta^3_\beta F_B^2\right)\,,
\nonumber
\\
&{R}^{\rho\sigma}_{\alpha\beta} = 
\delta^3_\alpha\delta^3_\beta \delta_3^\rho\delta_3^\sigma D_B^1 \!
+ \delta^3_\alpha\delta^4_\beta\delta_3^\rho\delta_4^\sigma
D_B^2 \! + \delta^3_\alpha\delta^4_\beta\delta_4^\rho\delta_3^\sigma
D_B^3 \!  +  \delta^4_\alpha\delta^3_\beta\delta_4^\rho\delta_3^\sigma
D_B^4 \! + \delta^4_\alpha\delta^3_\beta\delta_3^\rho\delta_4^\sigma
D_B^5 \!+
\delta^4_\alpha\delta^4_\beta \delta_4^\rho\delta_4^\sigma D_B^6 ,
\nonumber
\\
&{R}^{\beta b}_{a\alpha} = 
 \delta_a^b \delta^\beta_\alpha G^\alpha_B
\qquad
{R}^{b\beta}_{a\alpha} =  
 \delta_a^b \delta^\beta_\alpha H^\alpha_B\,,
 \nonumber
 \\
&{R}^{\beta b}_{\alpha a} =  
 \delta_a^b \delta^\beta_\alpha K^\alpha_B
\qquad
{R}^{b\beta}_{\alpha a} = \delta_a^b \delta^\beta_\alpha L^\alpha_B
\label{R-matrix_with_dof_result}
\end{align}
where, for the sake of notational simplicity, we have omitted the $(x^+,x^-)$ dependence in all the reflection functions.

Commutation with the fermionic generators $\mathbb{Q}_3^{\ a}$ and $\mathbb{S}_a^{\ 3}$ further constrains the reflection functions, but three out of twenty remain undetermined. This means there is a lot of freedom to construct a reflection matrix that preserves the $SU(1|2)$ symmetry. If one naively fixes the undetermined functions arbitrarily, the BYBE would typically not be satisfied. Given that the original singlet boundary was integrable, it seems natural to restrict these undetermined functions so that the new reflection matrix \eqref{Rdof} is also that of an integrable boundary. We shall complete the determination of the reflection matrix by imposing Yangian symmetry in the next subsection.

\subsection{Boundary Yangian symmetry} 
\label{boundary yangian}

The Yangian $Y({\bf g})$ is an infinite dimensional extension of a simple Lie algebra ${\bf g}$ introduced by Drinfeld \cite{Drinfeld:1985rx} and it is typically associated  with integrable models. In the context of AdS/CFT scattering problems, it was early recognized that the $su(2|2)$ symmetry admits a Yangian extension $Y(su(2|2))$ \cite{Beisert:2006fmy} and  has played a crucial role in the construction of the S-matrix for excitations in general multi-particle representations \cite{Arutyunov:2009mi}. Specifically, the Yangian extension of a bulk Lie algebra $\mathbf g$ is a deformation of the universal enveloping algebra of the polynomial $\mathbf g(u)$. It is generated by grade-$0$ generators $\mathbb J^A$ of $\mathbf g$ and grade-$1$ generators $\hat{\mathbb J}^A$ of $Y(\mathbf g)$. Their commutators have the form
\begin{equation}
    [\mathbb J^A,\mathbb J^B]=f^{AB}_{\;\;\;\;\;C}\;\mathbb J^C,\;\;\;\;[\mathbb J^A,\hat{\mathbb J}^B]=f^{AB}_{\;\;\;\;\;C}\;\hat{\mathbb J}^C,
\end{equation}
and must obey the Jacobi and Serre relations which, following \cite{Beisert:2006fmy}, are realized as follows,
\begin{equation}
    \begin{aligned}
& {\left[\mathbb{J}^{[A},\left[\mathbb{J}^B, \mathbb{J}^{C]}\right]\right]=0, \quad\left[\mathbb{J}^{[A},\left[\mathbb{J}^B, \hat{\mathbb{J}}^{C]}\right]\right]=0,} \\
& {\left[\hat{\mathbb{J}}^{[A},\left[\hat{\mathbb{J}}^B, \mathbb{J}^{C]}\right]\right]=\frac{1}{4} f_{\;\;\;\;\; D}^{AG} f_{\;\;\;\;\; E}^{B H} f_{\;\;\;\;\; F}^{C K} f_{G H K} \mathbb{J}^{\{D} \mathbb{J}^E \mathbb{J}^{F\}} .}
\end{aligned}
\end{equation}

The Yangian is a Hopf algebra and its co-products provide the rule for the action of the algebra generators on a tensor product of two  representations. The co-products of the grade-$0$ and grade-$1$ generators take the form
\begin{equation}
    \Delta \mathbb{J}^A=\mathbb{J}^A \otimes 1+1 \otimes \mathbb{J}^A, \quad \Delta \hat{\mathbb{J}}^A=\hat{\mathbb{J}}^A \otimes 1+1 \otimes \hat{\mathbb{J}}^A+\frac{1}{2} 
    f_{\ \,BC}^A \mathbb{J}^B \otimes \mathbb{J}^C .
\end{equation}

When constructing finite dimensional representations of $Y(\mathbf g)$ one has to consider the ``evaluation representation'' \cite{Beisert:2006fmy}, where the grade-$1$ generators take the form 
\begin{equation}
    \hat{\mathbb J}^A\ket{u}=ig\,
    u\,\mathbb J^A\ket{u}.
\end{equation}
For bulk magnons $u$ is a known function related to the rapidity \cite{Beisert:2006fmy}, namely 
 
\begin{equation}
    u=\frac{1}{2}\left(x^++\frac{1}{x^+}+x^-+\frac{1}{x^-}\right).
\end{equation}
For the boundary degree of freedom, we will parametrize the action of the grade-$1$ generators as
\begin{equation}
    \hat{\mathbb J}^A\ket{u_{B}}=u_B\;\mathbb J^A\ket{u_B},
\end{equation}
where  $u_B$ will be fixed when imposing Yangian symmetry.

When the bulk theory is restricted by a boundary only a remnant $Y(\mathbf h,\mathbf g)$ of the bulk Yangian symmetry survives, where $\mathbf h$ is the Lie subalgebra preserved by the boundary. As discussed in \cite{MacKay:2010ey,MacKay:2010zb,Palla:2011eu,Ahn:2010xa} $(\mathbf h,\mathbf g)$ must form a symmetric pair
\begin{equation}
    \mathbf{g}=\mathbf{h}+\mathbf{m}, \quad[\mathbf{h}, \mathbf{h}] \subset \mathbf{h}, \quad[\mathbf{h}, \mathbf{m}] \subset \mathbf{m}, \quad[\mathbf{m}, \mathbf{m}] \subset \mathbf{h}.
\end{equation}
 In what follows, indices $i,j,k$ run over $\mathbf h$ and the  indices $p,q,r$ run over $\mathbf m$. The remnant $Y(\mathbf h,\mathbf g)$ is generated by grade-$0$ generators $\mathbb J^i$ and   twisted grade-$1$ generators $\tilde{\mathbb J}^p$ where
\begin{equation}
    \tilde{\mathbb J}^p=\hat{\mathbb J}^p+\frac{1}{2}f^p_{\ qi}\mathbb J^q\mathbb J^i.
\end{equation}
 The co-products of these twisted generators are given by 
 \begin{equation}
\begin{aligned}
\Delta \tilde{\mathbb{J}}^p= &\  \Delta \hat{\mathbb{J}}^p+\frac{1}{2} f_{\ q i}^p \Delta \mathbb{J}^q \Delta \mathbb{J}^i 
\\
= &\  \tilde{\mathbb{J}}^p \otimes 1+1 \otimes \tilde{\mathbb{J}}^p+f_{\ q i}^p \mathbb{J}^q \otimes \mathbb{J}^i.
\end{aligned}
\label{coproducts}
 \end{equation}
In the problem we are focused on, $\mathbf g$ is $su(2|2)$ and $\mathbf h$ is $su(1|2)$, whose generators are $\mathbb R_a^{\;b}, \mathbb L_3^{\;3}=-\mathbb L_4^{\;4}, \mathbb Q_3^{\;a},\mathbb S_a^3$ and $\mathbb H$.

We will demand that the remnant $Y(\mathbf h,\mathbf g)$ is preserved during the boundary scattering process. As the reflection matrix acts on a tensor product of two representations \eqref{Rdof}, it must therefore commute with the co-products of the $Y(\mathbf h,\mathbf g)$ generators. We first consider the commutation with the co-products of grade-0 generators $\Delta \mathbb J^i$,
\begin{equation}
[R,\Delta \mathbb J^i]=[R, \mathbb J^i\otimes 1+1\otimes \mathbb J^i]=0\,.
\end{equation}
This leads precisely to the matrix structure given in \eqref{R-matrix_with_dof_result}. Thus, it remains to impose the commutation with the twisted grade-$1$ generators co-products \eqref{coproducts}.  The effect of this imposition would be to further constrain the functions appearing in \eqref{R-matrix_with_dof_result}.
\begin{align}
\Delta\tilde{\mathbb{C}} = & \ \left(\hat{\mathbb{C}}+\tfrac{1}{2}\mathbb{C}\,\mathbb{H}\right)\otimes1+1\otimes\left(\hat{\mathbb{C}}+\tfrac{1}{2}\mathbb{C}\,\mathbb{H}\right)+\mathbb{C}\otimes\mathbb{H},\nonumber \\
\Delta\mathbb{\tilde{C}}^{\dagger} = & \ \left(\hat{\mathbb{C}}^{\dagger}-\tfrac{1}{2}\mathbb{C}^{\dagger}\,\mathbb{H}\right)\otimes1+1\otimes\left(\hat{\mathbb{C}}^{\dagger}-\tfrac{1}{2}\mathbb{C}^{\dagger}\,\mathbb{H}\right)-\mathbb{C}^{\dagger}\otimes\mathbb{H}\,.
\end{align}
The vanishing of the commutators
$[R,\Delta\tilde{\mathbb{C}}]=[R, \Delta\mathbb{\tilde{C}^{\dagger}} ]=0$
gives rise to two conditions for the boundary algebra parameters
\begin{equation}
  \begin{aligned}
(\eta_1+2u_B\eta_1+2i\eta_2)\eta_2+2ig(\eta_1+2i\eta_2)\eta_3= &\  0\,,\\
   i(2u_B-1) \eta_2^2 \eta_3 + g \eta_2 \bigl( 2i \eta_1 \eta_2 + (3 - 2u_B) \eta_3^2 \bigr) + g^2 \bigl(2i \eta_3^3  -4 \eta_1 \eta_2 \eta_3 \bigr)= &\  0\,.\\
  \end{aligned}
  \label{algebrapara}
\end{equation}
These equations can be solved,  for example, by expressing $u_B$ and $\eta_1$ in terms of $\eta_2$ and $\eta_3$.
After solving these two constraints, the commutation with 
\begin{align}
\Delta\tilde{\mathbb{L}}_{3}^{\ 4}  
= &\  
\left(\hat{\mathbb{L}}_{3}^{\ 4}
+\mathbb{L}_{3}^{\ 4}\,\mathbb{L}_{3}^{\ 3}+\tfrac{1}{2}\mathbb{S}_{c}^{\ 4}\,\mathbb{Q}_{3}^{\  c}\right)\otimes1
+1\otimes\left(\hat{\mathbb{L}}_{3}^{\ 4}
+\mathbb{L}_{3}^{\ 4}\,\mathbb{L}_{3}^{\ 3}+\tfrac{1}{2}\mathbb{S}_{c}^{\ 4}\,\mathbb{Q}_{3}^{\  c}\right)\nonumber 
\\
& +2\mathbb{L}_{3}^{\ 4}\otimes\mathbb{L}_{3}^{\ 3}+\mathbb{S}_{c}^{\ 4}\otimes\mathbb{Q}_{3}^{\  c}\,,
\end{align}
fixes the remaining matrix elements in terms of a single overall function. A list of the functions appearing in the reflection matrix is given in appendix \ref{lista}, where all of them are written  in terms of $\eta_2$ and $\kappa$, with $\eta_3=-i\kappa\;\eta_2$. 
No additional constraints are obtained from the commutation with the remaining co-products. We have checked that this reflection matrix satisfies the BYBE. Therefore, we have found a  family of solutions to the BYBE with two parameters. We should clarify the physical meaning of them. 

In the case of $\kappa$, it is directly related to 
the energy of the boundary degree of freedom by
\begin{equation}
    E_B=
    \frac12({\sf a}_B{\sf d}_B+{\sf b}_B{\sf c}_B) =  
    \frac{1}{2}+ g \kappa\,.
    \label{Energykappa}
\end{equation}
In contrast, although we keep it for completeness of the mathematical solution to the BYBE,
$\eta_2$ is not physically relevant. First, it does not affect the energy or other central charges of the boundary degree of freedom. Second, although some components of the reflection matrix do depend on it, $\eta_2$ does not enter the Bethe-Yang equations, ensuring the spectrum is independent of this parameter.  For example, when all the magnons in the state have the same flavour, one of $A_B$, $D^1_B$ or $D^6_B$ will appear in the Bethe-Yang equations, and they are all independent of $\eta_2$. For a more general state, the formulation of a nested Bethe ansatz would be necessary.  Although this is beyond the scope of the present article, a simple analysis of how the reflection matrix components scale with $\eta_2$ shows that level II reflection factors will also be independent of $\eta_2$\footnote{To determine level II reflection factors $R_{II}^\alpha(y)$, one would need to solve compatibility conditions similar to those in eqs. (41) and (42) of \cite{Correa:2009dm}. Since $H^\alpha_B$ and $K^\alpha_B$ are independent of $\eta_2$ and $L^\alpha_B\propto \eta_2$ while $G^\alpha_B\propto 1/\eta_2$, the compatibility conditions would be solved with $R_{II}^\alpha(y)$ independent of $\eta_2$.}. Another way to see that $\eta_2$ is unphysical is to note it appears as a multiplicative factor in certain specific components of the reflection matrix: when the boundary degree of freedom changes from bosonic to fermionic the reflection amplitude is proportional to $\eta_2$, whereas the opposite process is proportional to $1/\eta_2$. Thus, these factors can be simply removed by redefining the relative normalization between the bosonic and fermionic states at the boundary.

Let us now focus on our concrete realization of  a magnon at the boundary of an ABJM Wilson loop spin chain. The binding energy of such boundary degree of freedom  is \eqref{exactenergypole}, and to reproduce it with \eqref {Energykappa}, the choice of $\kappa$ should be
\begin{equation}
    \kappa=1+\frac{i}{x_B}\,,
    \label{espefickapppa0}
\end{equation}
where $x_B$ is defined in \eqref{expaQ1bbs}. Therefore, the reflection matrix in this particular case can be explicitly written by taking the expressions in appendix \ref{lista} for this specific value of $\kappa$.

 Since the boundary degree of freedom in this concrete example originates as a boundary bound state, it should therefore be possible to obtain the same reflection matrix via a {\it boundary bound state bootstrap mechanism} as the one described in \cite{Ghoshal:1993tm}, which is schematically depicted in Fig. \ref{fig: fusion} and would impose the equation
\begin{equation}
g_{A\,0}^{\ \, B}\, R_{C \,B}^{C' \hspace{0.02cm} B' }(x^\pm) 
=  
g_{G\,0}^{\ \, B'}\,S_{C\, A}^{D\,E}(x^\pm,x_*^\pm) \, R_{D}^{F}(x^\pm)\,  S_{E \, F}^{G \, C'}(x_*^\pm,-x^\mp) \,,
\label{bootstrapformula}
\end{equation}
where $x_*^+=x_B$ and $x_*^-=-i$, while $g_{A\,0}^{\ \, B}$ are the particle-boundary coupling constants that can be read from the residue of the pole in the singlet boundary reflection matrix. Since  the bound state is formed with a singlet boundary, the energy and charges of the incoming particle $A$ are the same as those of the trapped particle $B$. In particular,
\begin{equation}
    g_{A\,0}^{\ \, B} = \delta_A^B f_A\,,
\end{equation}
and eq. \eqref{bootstrapformula} becomes
\begin{equation}
R_{C \,A}^{C' \hspace{0.02cm} B' }(x^\pm) 
=  
\frac{f_{B'}}{f_A} S_{C\, A}^{D\,E}(x^\pm, x_*^\pm) \, R_{D}^{F}(x^\pm)\,  S_{E \, F}^{B' \hspace{0.02cm} C'}(x_*^\pm,-x^\mp)\,.
\label{bootstrapformula2}
\end{equation}
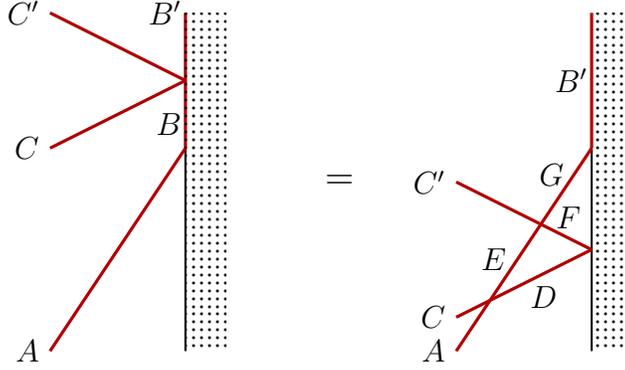
\begin{figure}[h!]
\begin{center}
\begin{tikzpicture}[scale=0.9]
\draw[thick] (0,1) -- (0,4);
\draw[very thick,red!70!black] (0,4) -- (0,6);
\draw[very thick,red!70!black] (0,4) -- (-2,1);
\draw[very thick,red!70!black] (-2,4) -- (0,5);
\draw[very thick,red!70!black] (-2,6) -- (0,5);
\fill[pattern=dots, pattern color=black] (0,6) rectangle (0.6,1);
\node[left] at (-2,6) {$C'$};
\node[left] at (-2,1) {$A$};
\node[left] at (-2,4) {$C$};
\node[left] at (0.1,4.35) {$B$};
\node[left] at (0.1,6) {$B'$};
\node[left] at (2.65,3.5) {\bf \large $=$};
\draw[thick] (6,1) -- (6,4);
\draw[very thick,red!70!black] (6,4) -- (6,6);
\draw[very thick,red!70!black] (6,4) -- (4,1);
\draw[very thick,red!70!black] (4,1.5) -- (6,2.5);
\draw[very thick,red!70!black] (4,3.5) -- (6,2.5);
\fill[pattern=dots, pattern color=black] (6,6) rectangle (6.6,1);
\node[left] at (4,1.5) {$C$};
\node[left] at (4,1) {$A$};
\node[left] at (4,3.5) {$C'$};
\node[left] at (4.9,2.36) {$E$};
\node[left] at (6.1,5) {$B'$};
\node[left] at (5.65,1.8) {$D$};
\node[left] at (6.,2.98) {$F$};
 (2∣2) representation with bosonic singlets and a fermionic doublet i\node[left] at (5.75,3.6) {$G$};
\end{tikzpicture}
\end{center}
\caption{Boundary bound state bootstrap condition.}
\label{fig: fusion}
\end{figure}

By changing the relative normalization of the magnons at the boundary one can absorb the ${f_{B'}}$ and ${f_A}$ factors. Thus,  eq. \eqref{bootstrapformula2} becomes in that case
\begin{equation}
R_{C \,A}^{C' \hspace{0.02cm} B' }(x^\pm) 
=  S_{C\, A}^{D\,E}(x^\pm,x_*^\pm) \,
R_{D}^{F}(x^\pm)\,  S_{E \, F}^{B' \, C'}(x_*^\pm,-x^\mp)\,.
\label{bootstrapformula3}
\end{equation}
It is straightforward to check that either \eqref{bootstrapformula2} or \eqref{bootstrapformula3} satisfies unitarity, crossing symmetry, and the BYBE.

It might be instructive to write down some components explicitly and compare them with those
obtained by imposing Yangian symmetry,
given in appendix \ref{lista}. For example
\begin{align}
\frac{R _{1\,1}^{1\,1}(x^\pm)}{R_{1\,4}^{4\,1}(x^\pm)}  = &\  \frac{A(x^\pm,x_*^\pm)  A(x_*^\pm,-x^\mp)}{G(x^\pm, x_*^\pm)  K(x_*^\pm,-x^\mp)-
\sqrt{\frac{x^+}{x^-}} H(x^\pm,x_*^\pm)  G(x_*^\pm,-x^\mp)}  \nonumber
\\
= &\ \sqrt{\frac{x^-}{x^+}} \frac{x^+(x^--i)(x^--x_B)}{x_B(x_B+i)(x^++x^-)}\frac{\eta(\zeta \tfrac{x^+}{x^-},x_B,-i)}{\eta(\zeta,x^+,x^-)}\,.
\label{simplerratio}
\end{align}
This ratio should agree (up to the unphysical $\eta_2$) with 
\begin{equation}
\frac{A_B}{G_B} =  \frac{x^+ - x^-(1 + x^{+2}) +  2ix^{-3}x^+\kappa + x^{-2}\bigl(x^+ -i \kappa -i x^+(x^+ +i \kappa)\kappa\bigr)}
{ x^- (x^- + x^+) ( \kappa+\tfrac{1}{g} ) }
\frac{\sqrt{f}\eta_2}{\eta} \,,
\label{AoverG}
\end{equation}
When evaluated at $\kappa = 1 + i/x_B$, this expression, despite its dissimilar appearance, coincides with \eqref{simplerratio} after using the spectral parameter constraints. In the same fashion, we have checked that the entire list of functions determined by Yangian symmetry coincides with the
expressions $-$typically simpler$-$ computed  from the bootstrap formula \eqref{bootstrapformula3}.

In the next section, we will compare some of these results 
with explicit weak coupling computations. In order to test the matrix structure, it is useful to consider some specific ratios of components, such as
\begin{align}
\frac{{R}_{1\, 2}^{2\, 1}}
 {R_{1\, 1}^{1\, 1}} = 
 &\ \frac{2i (x_B + g(i + x_B)) (x^+ - x^-) (x^- + x^+) (1 + x^{+2})}
{g (x_B - x^-) (-i + x^-) x^{+2} \bigl(-1 + x_B (i - x^- + x^+)\bigr)}
\nonumber\\
=  &\
\frac{2(1-e^{2ip})}{2-e^{ip}}
+ \lambda \frac{2(1-e^{2ip})(1-e^{ip})}{2-e^{ip}} + {\cal O}(\lambda^2)\,.
\label{ratioABAexpa}
\end{align}

We conclude this section by noting that the reflection matrix determined by Yangian symmetry and via the boundary bound state bootstrap condition coincides not only for our concrete example \label{espefickapppa} but also for more general values of $\kappa$. Indeed, we can write a more general bootstrap equation
\begin{equation}
R_{C \,A}^{C' \hspace{0.02cm} B' }(x^\pm) 
=  S_{C\, A}^{D\,E}(x^\pm,y_B^\pm) \,
R_{D}^{F}(x^\pm)\,  S_{E \, F}^{B' \, C'}(y_B^\pm,-x^\mp)\,,
\label{bootstrapformula4}
\end{equation}
where the spectral parameters $y_B^\pm$ are chosen to reproduce the binding energy \eqref{Energykappa},
\begin{equation}
 y_B^\pm =  \frac{i}{2g\kappa} \left(
 \sqrt{\kappa (g\kappa+1)(g(\kappa^2-4)+\kappa)}\pm \kappa(g\kappa+1)
 \right)\,.
 \label{yBs}
\end{equation}
Thus, in the generic case the ratio we have considered above becomes
\begin{equation}
\frac{R _{1\,1}^{1\,1}(x^\pm)}{R_{1\,4}^{4\,1}(x^\pm)}  
=
\frac{x^+(x^-+y_B^-)(x^--y_B^+)}{y_B^+(y_B^+-y_B^-)(x^++x^-)}\frac{\eta(\zeta \tfrac{x^+}{x^-},y_B^+,y_B^-)}{\eta(\zeta,x^+,x^-)}\,,
\label{simplerratiogeneric}
\end{equation}
which is just another way of expressing \eqref{AoverG}.

The use of the boundary bound state bootstrap mechanism to obtain the reflection matrix for a generic binding energy makes sense as long as the spectral parameters \eqref{yBs} can be explained by appropriate poles and zeros in the overall dressing factor of the singlet boundary reflection matrix. Such poles and zeros can always be introduced with a CDD factor \cite{Bajnok:2004jd}. In the present case, it can take the form,
\begin{equation}
 R_{\sf CDD}(x^\pm) 
 =\frac{y^+_B-x^+}{y^+_B+x^-} 
 \frac{y^+_B+\frac1{x^-}}{y^+_B-\frac1{x^+}}\,.
\end{equation}
Another requirement to describe a boundary bound state is that the imaginary part of the momentum associated with \eqref{yBs} must be  negative, which is always the case as long as 
\begin{equation}
   |E_B| = |\tfrac{1}{2} +g\kappa| > \tfrac{1}{2}
   \sqrt{1+16g^2} = E(\pi)\,.
   \label{rango}
\end{equation}

\section{Weak coupling verifications}
\label{weak}

As already discussed, the spectral problem for operators inserted into the 1/2 BPS Wilson line, provides a concrete realization for a spin chain with $SU(1|2)$ symmetry that can allocate boundary degrees of freedom. Thus, explicit computations of the scaling dimensions for such operators
can be used to test our proposal for the reflection matrix in the weak coupling limit.

We begin by considering the so-called $SU(2)$ sector for operators inserted into the 1/2 BPS Wilson line. This means that we will consider only certain components of type $A$ magnons, {\it i.e.}  we restrict our attention to insertions in which some of the $C_1$ fields in the reference state \eqref{susyinsertion0} are replaced by $C_3$. The Hamiltonian giving the anomalous dimension of these inserted operators was computed in \cite{Correa:2023lsm}
\begin{equation}
\label{weakcouplinghamitonian0}
{\bf H}= ( \lambda + \lambda^2 \, \beta_0 ) {\bf V}_0 + \lambda^2 \sum_{n=0}^{\infty} ({\bf 1}-{\bf P}_{n,n+1}) \,.
\end{equation}
Here we have considered a semi-infinite open chain, as we are primarily interested in computing reflection factors. The order $\lambda^2$ term is the typical nearest neighbour interaction of a Heisenberg spin chain, while the order $\lambda$ term describes a boundary interaction. The operator ${\bf V}_0$ yields 0 or 1 when the leftmost site of the chain is occupied by $C_1$ or $C_3$, respectively. 

To begin with, we would like to fix $\beta_0$, the contribution of ${\bf V}_0$ at order $\lambda^2$ that was left undetermined in \cite{Correa:2023lsm}. We can infer the value of $\beta_0$ from some two-loop scaling dimension computation. For instance,  consider the operator insertion
\be 
{\cal O}_3 = 
\left(
\begin{array}{cc}
 0 &  C_3
\\
0 &  0
\end{array}\right)\,,
\label{insertionC3}
\ee
whose scaling dimension should be
\begin{equation}
 \Delta_{{\cal O}_3} = \frac12 + 2 \lambda + 2 \lambda^2 \beta_0\,. 
\end{equation}
 Its explicit computation would involve a significant number of two-loop Feynman diagrams. However, we can obtain the same answer via a shortcut using known results in the literature.

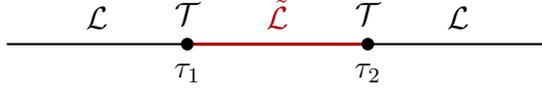
\begin{figure}[h!]
\begin{center}
\begin{tikzpicture}[scale=1.2]
\draw[thick] (0,0) -- (6,0);
\draw[very thick,red!70!black] (2,0) -- (4,0);
\fill (2,0) circle (2pt);
\fill (4,0) circle (2pt);
\node[below] at (2,-0.1) {$\tau_1$};
\node[below] at (4,-0.1) {$\tau_2$};
\node[above]  at (1,0.05) {$\mathcal{L}$};
\node[above, black]  at (2,0.1) {$\mathcal{T}$};
\node[above, red!70!black] at (3,0.05) {$\tilde{\mathcal{L}}$};
\node[above, black]   at (4,0.1) {$\mathcal{T}$};
\node[above]  at (5,0.05) {$\mathcal{L}$};
\end{tikzpicture}
\end{center}
\caption{Insertion of two constant primary operators. 
They work as DCOs, transforming the connection from $\mathcal{L}$ to $\tilde{\mathcal{L}}$ and back, where the change is simply a sign in the off-diagonal fermions.}
\label{TauDCOs}
\end{figure}

First, note that the operator \eqref{insertionC3} is a descendant of a constant primary operator ${\cal T}$ \cite{Gorini:2022jws} (see appendix \ref{app: review WL}). Consequently, they share the same anomalous dimension. The anomalous dimension of  ${\cal T}$ is deduced from a 2-point correlator of the form \eqref{2ptintheline}. Given that ${\cal T} = {\rm diag}(\mathds{1}_N,-\mathds{1}_N)$,
when two such constant operators are inserted at $\tau_1$ and $\tau_2$, their net effect is changing the sign of the off-diagonal components of the Wilson line for the interval $\tau_1<\tau<\tau_2$, which in turn amounts to a change of sign of $\eta^\alpha_I$ and $\bar\eta_\alpha^I$, the coupling to the fermionic fields in the superconnection (see figure \ref{TauDCOs}). Thus, the constant primary operator ${\cal T}$ is merely a defect changing operator (DCO) \cite{Kim:2017sju} \footnote{We thank Nadav Drukker for a discussion about the relation between ${\cal T}$ and defect changing operators.}.  Interestingly, such a change of sign in the fermionic fields in the superconnection can be interpreted as a geometrical cusp angle $\varphi$ in the limit $\varphi \to 2\pi$.  While a cusp angle of $2\pi$ would be trivial for a bosonic Wilson line, it is not for a fermionic one, as it changes the signs of fermionic fields.  Thus, inserting two ${\cal T}$ operators is the same as inserting two $\varphi  \to 2\pi$ cusps. The anomalous dimension of the latter is obtained from the expectation value of the cusped Wilson line (when the cusp at $\tau_2$ is sent to $+\infty$).

Fortunately, the cusp anomalous dimension for Wilson loops in the ABJM model has already been computed up to two-loop order in \cite{Griguolo:2012iq},
\begin{equation}
\Gamma_{\rm cusp} (\varphi) 
= -\lambda \left(\frac{1}{\cos\tfrac{\varphi}{2}}-1\right)
-\frac{\lambda^2}{2} \log(\cos\tfrac{\varphi}{2})^2 
+{\cal O}(\lambda^3)\,.
\end{equation}
Since,
\begin{equation}
    \Delta_{{\cal O}_3} -\frac12 = \lim_{\varphi\to 2\pi} \Gamma_{\rm cusp} (\varphi) = 2\lambda + {\cal O}(\lambda^3)\,,
\end{equation}
we conclude that $\beta_0 = 0$. Thus, the Hamiltonian \eqref{weakcouplinghamitonian0} becomes
\begin{equation}
\label{weakcouplinghamitonian}
{\bf H}= \lambda {\bf V}_0 + \lambda^2 \sum_{n=0}^{\infty} ({\bf 1}-{\bf P}_{n,n+1}) \,.
\end{equation}

We now turn to study the eigenvalue problem of the Hamiltonian \eqref{weakcouplinghamitonian}. The diagonalization for states with a single magnon impurity was already done in \cite{Correa:2023lsm} by treating the bulk term of the Hamiltonian as a perturbation of the boundary term. The same result can be obtained if one regards the Hamiltonian \eqref{weakcouplinghamitonian} as an exact function of $\lambda$ and takes the small $\lambda$ limit at the end.

In a standard coordinate Bethe ansatz, the following wave-function can be proposed for a single magnon state
\begin{equation}
    |\psi(p) \rangle = \sum_{n=0} \left(e^{ip n} + R(p) e^{-ip n}\right) |n\rangle \,,
\label{onemagnonbethestate}
\end{equation}
where  $|n\rangle $ is a spin chain state with a single impurity in the $n^{\rm th}$ site. It is straightforward to check that \eqref{onemagnonbethestate} is an eigenstate of the Hamiltonian \eqref{weakcouplinghamitonian} with energy $4\lambda^2\sin^2\frac{p}{2}$, provided that
\begin{equation}
    R (p) = -\frac{1+ \lambda (e^{-ip}-1)}{1+ \lambda (e^{ip}-1)} \,.
    \label{Rnonper}
\end{equation}
In the weak coupling limit we recover $R (p) =-1$,  which is in agreement with the weak-coupling of the all-loop result \eqref{crossing ansatz}, as previously verified in \cite{Correa:2023lsm}. Beyond the leading weak-coupling order, the Bethe ansatz is asymptotic. Thus, to test higher order terms of the of the all-loop reflection factor, the ansatz \eqref{onemagnonbethestate} should incorporate contact terms. Additionally, including $\lambda^3$ terms into the Hamiltonian \eqref{weakcouplinghamitonian} would be necessary to compute the next to leading order reflection factor.

 The eigenstate with energy of order $\lambda$ appears as a pole in the reflection factor \eqref{Rnonper}. Indeed, for a complex momentum such that
\begin{equation}
    e^{ip_0} = 1-\frac{1}{\lambda}\,,
\end{equation}
the wave-function becomes
\begin{equation}
    |B_1\rangle = |\psi(p_0) \rangle = |0\rangle -\lambda |1\rangle + {\mathcal O}(\lambda^2)\,,
\end{equation}
and exhibits an energy
\begin{equation}
    E_B^{(1)} = \lambda + \lambda^2 +{\mathcal O}(\lambda^3)\,,
\end{equation}
which matches precisely the first two orders in the weak coupling expansion of the exact result \eqref{expaQ1bbs}. The matching of the order $\lambda^2$ confirms the value $\beta_0 =0$.

The main motivation of this section is to verify the reflection matrix in the case of a boundary degree of freedom in the weak coupling limit. For that, we need to proceed and consider states with more magnon
excitations, being at least one of them trapped at the boundary.
As in the case of the 1-magnon boundary bound state, the components of the corresponding wave-function decrease by powers of $\lambda$ as the leftmost magnon increases its distance from the boundary.  Up to the order we are working with and for states with $Q$ excitations, we consider wave functions of the form
\begin{equation}
    |\psi\rangle =  \sum_{n_i} a_{n_2,\cdots,n_{Q}} 
    |0,n_2,\cdots,n_{Q}\rangle 
    +\lambda 
    \sum_{n_i} b_{n_2,\cdots,n_{Q}} |1,n_2,\cdots,n_{Q}\rangle  + {\mathcal O}(\lambda^2)\,,
    \label{genericansatz}
\end{equation}
where $|n_1,n_2,\cdots,n_{Q}\rangle$ is a state with impurities in the $n_i^{\rm th}$ sites. Demanding \eqref{genericansatz} is an eigenstate of ${\bf H}$ with eigenvalue
$E = \lambda + e_2 \lambda^2 + {\mathcal O}(\lambda^3)$, immediately fixes that
\begin{equation}
b_{n_2,\cdots,n_{Q}}= -a_{n_2,\cdots,n_{Q}}\,.
\end{equation}

Let us turn to some concrete examples. First, we would like to verify the existence of boundary states with two magnons trapped at the boundary, whose exact energy is given by  \eqref{exactenergypoleQ}  with  $Q = 2$. The ansatz for the wave function is in this case   
\begin{equation}
    |B_2\rangle =  \sum_{n=1} a_n |0,n\rangle -\lambda \sum_{n=2} a_n |1,n\rangle  + {\mathcal O}(\lambda^2)\,.
    \label{B2ansatz}
\end{equation} 
Demanding \eqref{B2ansatz} is an eigenstate of \eqref{weakcouplinghamitonian} with eigenvalue
$\lambda+ e_2 \lambda^2$, we obtain a series of conditions on the coefficients, namely 
\begin{equation}
    a_1 = 1\,,\quad  
    a_2 = 1-e_2\,,\quad
    a_{n+1}=(3-e_2)a_n -a_{n-1}\,.
\end{equation}
The first condition is just an arbitrary normalization. We are left with a second order recurrence equation, whose solution can be put in terms of an arbitrary value of $e_2$,
\begin{align}
a_n =  &  \left(\frac{2-e_2}{\sqrt{(e_2-5) (e_2-1)}}+1\right) \left(\frac{3-e_2-\sqrt{(e_2-5)
   (e_2-1)}}{2}\right)^n \nonumber 
   \\
 &   -\left(\frac{2-e_2}{\sqrt{(e_2-5) (e_2-1)}}-1\right)
\left(
\frac{3-e_2+\sqrt{(e_2-5)(e_2-1)}}{2}\right)^n\,.
\end{align}
The final constraint, in order to have a normalizable wave-function, is that $\lim_{n\to\infty} a_n = 0$, which is only possible for
\begin{equation}
e_2 = \frac12\,, \quad \Rightarrow \quad a_n = 2^{1-n}\,.
\end{equation}
Therefore,
\begin{align}
    |B_2\rangle = &\  \sum_{n=1} (\tfrac{1}{2})^{n-1} |0,n\rangle -\lambda \sum_{n=2} (\tfrac{1}{2})^{n-1}  |1,n\rangle  + {\mathcal O}(\lambda^2)\,,
    \\
    E_0^{(2)} =  &\ \lambda + \frac{\lambda^2}2 +{\mathcal O}(\lambda^3)\,.
\end{align}

This eigenstate, whose wave-function describes two magnons trapped at the boundary, possesses an energy that matches the first two orders in the weak coupling expansion of the exact result \eqref{exactenergypoleQ} when  $Q = 2$.

Of course, larger bound states can also be trapped at the boundary, and evidence of that can also be seen with the Hamiltonian \eqref{weakcouplinghamitonian}. 
By acting explicitly with the Hamiltonian \eqref{weakcouplinghamitonian} on 
\begin{align}
    |B_Q\rangle = & \! \sum_{0<n_i< n_{i+1}} \!\!(\tfrac{1}{2})^{n_2} (\tfrac{2}{3})^{n_3}
    \cdots (\tfrac{Q-1}{Q})^{n_Q} |0,n_2,\cdots  n_Q \rangle 
   \nonumber\\ 
    & \ -\lambda \!\! \sum_{1<n_i< n_{i+1}} \!\! (\tfrac{1}{2})^{n_2} (\tfrac{2}{3})^{n_3}
    \cdots (\tfrac{Q-1}{Q})^{n_Q} |1,n_2,\cdots  n_Q \rangle   + {\mathcal O}(\lambda^2)\,,
\end{align}
one can verify that it is an eigenstate with eigenvalue
\begin{align}
    E_0^{(Q)} =  &\ \lambda + \frac{\lambda^2}{Q} +{\mathcal O}(\lambda^3)\,,
\end{align}
also in  agreement with \eqref{exactenergypoleQ}, now for any $Q$.

The above computation serves as a verification of the expected poles in the overall dressing factors of the reflection matrices  presented in \ref{R-matrix vac}. We would also like to check the matrix structure of the reflection matrix derived in section \ref{R-matrix with d.o.f}, which would require moving to a larger sector of insertions. For instance, we can consider the $SU(3)$ sector for type $A$ magnons. This means we will now consider insertions in which some of the $C_1$ fields in the reference state \eqref{susyinsertion0} are replaced by either $C_3$ or $C_4$. The Hamiltonian in this larger sector is still given by \eqref{weakcouplinghamitonian}.

We will consider a wave-function for two magnons, one trapped at the left boundary and the other propagating with momentum $p$. In this larger sector, the wave-function involves a proper reflection matrix,
\begin{align}
    |\psi(p) \rangle_{ab} = &\ \sum_{n=1} \left(e^{ip n} \delta_a^c\delta_b^d +  e^{-ip n} R_{ab}^{cd}(p) \right)|0,n\rangle_{cd}
    \nonumber\\
    &\ -\lambda\sum_{n=2} \left(e^{ip n} \delta_a^c\delta_b^d  +  e^{-ip n} {R}_{ab}^{cd}(p)\right) |1,n\rangle_{cd}
    + {\mathcal O}(\lambda^2)\,,
\label{onemagnonbethestatebdrydof}
\end{align}
where ${ab}$ in $|m,n\rangle_{ab}$ indicates the flavour of the impurities at $m^{\rm th}$ and $n^{\rm th}$ positions respectively. We will demand that \eqref{onemagnonbethestatebdrydof} is an eigenstate of the Hamiltonian \eqref{weakcouplinghamitonian} with some energy $ \lambda + \lambda^2 e(p)$. The impurities admit two possible flavours and the reflection has to commute with the $su(2)$ rotations of them, which restricts the reflection matrix to be of the form
\begin{equation}
  {R}_{ab}^{cd}(p) = \frac{A_L(p)+B_L(p)}{2}   \delta_a^c\delta_b^d + \frac{A_L(p)-B_L(p)}{2}   \delta_a^d\delta_b^c\,.
\end{equation}
A rather direct computation shows that
\begin{equation}
    {\bf H} |\psi(p) \rangle_{ab} = ( \lambda + \lambda^2 e(p))|\psi(p) \rangle_{ab} +{\mathcal O}(\lambda^3)\,
\end{equation}
requires that 
\begin{align}
\lambda + \lambda^2 e(p) = & \ \lambda + \lambda^2 + 4 \lambda^2 \sin^2\tfrac{p}{2}\,,    
\\
A_L(p) =  -\frac{1-2 e^{ip}}{1-2 e^{-ip}}\,,
\quad &\ \quad
B_L(p) = -1\,.
\end{align}
The energy is that of a magnon trapped at the boundary plus the energy of a magnon of momentum $p$. 
For comparison with the results of the previous section, it is convenient to compute the ratio for right reflection factors
\begin{equation}
 \frac{{R_{\sf R}}_{1\, 2}^{2\, 1}}
 {{R_{\sf R}}_{1\, 1}^{1\, 1}} = 
 \frac{A_L(-p)-B_L(-p)}{2A_L(-p)} 
 = \frac{2(1-e^{2ip})}{2-e^{ip}}\,,
 \label{ratioABA}
\end{equation}
which is in perfect agreement with the leading order of \eqref{ratioABAexpa}.

\section{Discussion}
\label{discu}
The main result of this paper is the reflection matrix preserving an $SU(1|2)$ symmetry for
a boundary that allocates a certain type of degree of freedom. We have obtained it by 
symmetry considerations. We have worked out in detail the case in which these degrees of freedom transform in a 4-dimensional representation. A concrete realization of this setup is found when studying insertions in the 1/2 BPS Wilson loop in ABJM. The boundary degree of freedom appears in that case as a boundary bound state due to the presence of poles in the dressing phase of the reflection matrix from a singlet boundary. 

Imposing the boundary remnant of the Yangian symmetry we found a whole family of integrable reflection matrices labelled by a continuous parameter\footnote{The reflection matrix we obtained imposing Yangian symmetry also depended on a second parameter, but that one was unphysical as it could be removed with a simple renormalization of the states describing the boundary.} $\kappa$, which specifies the energy of the boundary degree of freedom. 
When $\kappa$ lies in the region given by \eqref{rango}, the boundary degree of freedom can be interpreted as a boundary bound state. In these cases, the reflection matrix 
coincides with the result obtained from a boundary bound state bootstrap procedure. 
The concrete ABJM Wilson loop realization is singled out for $\kappa=1+i/x_B$. We used that example to test our results against explicit perturbative computations. It would be interesting to find the dual holographic 
description of the boundary bound states in that case
in terms of open strings in $AdS_4 \times \mathbb{CP}^3$, in order to further test our results in the strong coupling regime. For the values of $\kappa$ outside the region given by \eqref{rango}, as for example $\kappa=0$, the boundary degree of freedom can no longer be interpreted as a boundary bound state. Nevertheless, the reflection matrices obtained for such values are still sensible solutions to the BYBE.
It would be interesting to explore  whether such other cases for generic values of $\kappa$ can also be realized, either in ABJM or other physical models.

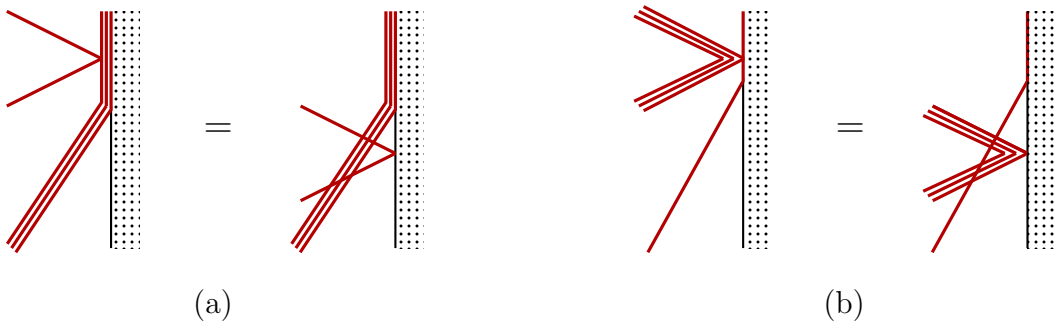
\begin{figure}[h!]
\begin{center}
\begin{tikzpicture}[scale=0.63]
\draw[thick] (0+0.1,1) -- (0+0.1,4);
\draw[very thick,red!70!black] (0-0.1,4+0.08) -- (0-0.1,6);
\draw[very thick,red!70!black] (0+0.1,4-0.1) -- (0+0.1,6);
\draw[very thick,red!70!black] (0,4) -- (0,6);
\draw[very thick,red!70!black] (0+0.09,4-0.09) -- (-2+0.09,1-0.09);
\draw[very thick,red!70!black] (0-0.09,4+0.09) -- (-2-0.09,1+0.09);
\draw[very thick,red!70!black] (0,4) -- (-2,1);
\draw[very thick,red!70!black] (-2.1,4) -- (-0.1,5);
\draw[very thick,red!70!black] (-2.1,6) -- (-0.1,5);
\fill[pattern=dots, pattern color=black] (0.14,6) rectangle (0.7,1);
\node[left] at (2.9,3.5) {\bf \large $=$};
\node[left] at (2.9,-0.2) {(a)};
\draw[thick] (6+0.1,1) -- (6+0.1,4);
\draw[very thick,red!70!black] (6-0.1,4+0.08) -- (6-0.1,6);
\draw[very thick,red!70!black] (6+0.1,4-0.1) -- (6+0.1,6);
\draw[very thick,red!70!black] (6,4) -- (6,6);
\draw[very thick,red!70!black] (6+0.09,4-0.09) -- (4+0.09,1-0.09);
\draw[very thick,red!70!black] (6-0.09,4+0.09) -- (4-0.09,1+0.09);
\draw[very thick,red!70!black] (6,4) -- (4,1);
\draw[very thick,red!70!black] (4.1,2) -- (6.1,3);
\draw[very thick,red!70!black] (4.1,4) -- (6.1,3);
\fill[pattern=dots, pattern color=black] (6.1,6) rectangle (6.7,1);
\end{tikzpicture}
\hspace{2.5cm}
\begin{tikzpicture}[scale=0.63]
\draw[thick] (0+0.1,1) -- (0+0.1,4.6);
\draw[very thick,red!70!black] (0+0.1,4.6-0.1) -- (0+0.1,6);
\draw[very thick,red!70!black] (0+0.09,4.6-0.09) -- (-2+0.09,1-0.09);
\draw[very thick,red!70!black] (-2.1,4) -- (-0.1,5);
\draw[very thick,red!70!black] (-2.1,6) -- (-0.1,5);
\draw[very thick,red!70!black] (-2.1+0.1,4-0.1) -- (-0.1+0.22,5);
\draw[very thick,red!70!black] (-2.1+0.1,6+0.1) -- (-0.1+0.22,5);
\draw[very thick,red!70!black] (-2.1-0.1,4+0.1) -- (-0.1-0.22,5);
\draw[very thick,red!70!black] (-2.1-0.1,6-0.1) -- (-0.1-0.22,5);
\fill[pattern=dots, pattern color=black] (0.14,6) rectangle (0.7,1);
\node[left] at (2.9,3.5) {\bf \large $=$};
\node[left] at (2.9,-0.2) {(b)};
\draw[thick] (6+0.1,1) -- (6+0.1,4.6);
\draw[very thick,red!70!black] (6+0.1,4.6-0.1) -- (6+0.1,6);
\draw[very thick,red!70!black] (6+0.09,4.6-0.09) -- (4+0.09,1-0.09);
\draw[very thick,red!70!black] (4.1,2) -- (6.1,3);
\draw[very thick,red!70!black] (4.1,4) -- (6.1,3);
\draw[very thick,red!70!black] (4.1-0.1,2+0.1) -- (6.1-0.24,3);
\draw[very thick,red!70!black] (4.1-0.1,4-0.1) -- (6.1-0.24,3);
\draw[very thick,red!70!black] (4.1,4) -- (6.1,3);
\draw[very thick,red!70!black] (4.1-0.2,2+0.2) -- (6.1-0.48,3);
\draw[very thick,red!70!black] (4.1-0.2,4-0.2) -- (6.1-0.48,3);
\fill[pattern=dots, pattern color=black] (6.1,6) rectangle (6.7,1);
\end{tikzpicture}
\end{center}
\vspace{-0.75cm}
\caption{Boundary bound state bootstrap with higher rank representations.
}
\label{bbsbQ}
\end{figure}

Our results provide the basis for future work on the physical interpretation of line defects in the ABJM model.
One could generalize our analysis by considering boundary degrees of freedom in higher-rank representations. The poles in the dressing phase for the reflection of magnons in totally symmetric representations \eqref{expaQQbbs} provide evidence of their existence. The reflection from these other boundaries can be obtained with a boundary bound state bootstrap procedure as sketched in Fig \ref{bbsbQ}(a).

Another possible generalization is to consider the reflection of higher-rank magnons from a boundary whose  impurity transforms in the fundamental representation.  The reflection matrix in these other cases can also be derived with the boundary bound state bootstrap mechanism, now depicted in Fig \ref{bbsbQ}(b). The reflection matrices presented in section \eqref{R-matrix vac} correspond to bound states in a totally symmetric representation, {\it i.e.} bound states of physical kinematics. It would also be very interesting to consider the case of totally anti-symmetric representations, describing bound states of the mirror theory. The reflection matrix for such $Q$-magnons, required for the bootstrap procedure in  Fig \ref{bbsbQ}(b), cannot be determined by the grade-0 generators alone. To derive it, one would need to impose Yangian symmetry, as done in \cite{Palla:2011eu}.  It is precisely this case which might play a significant role for the physical interpretation of line defects in the ABJM model. The reflection matrix of bound states of magnons in the mirror theory are needed for computing L\"uscher finite size corrections and for a TBA formulation. One concrete application of such  L\"uscher computations would be to determine the finite size correction to the scaling dimension of the operator ${\cal O}_3$ given in \eqref{insertionC3}. This operator belongs to the same multiplet as certain operators that drive supersymmetric deformations of the 1/2 BPS Wilson loop \cite{Ouyang:2015iza,Ouyang:2015bmy}.  Determining its scaling dimension would be crucial to clarify whether these deformations should be interpreted  as marginal deformations or RG flows on the conformal line defect \cite{Correa:2019rdk,Castiglioni:2022yes,Castiglioni:2023uus}.

\subsection*{Acknowledgements}

We would like to thank N. Drukker, M. Lagares, V. Regelskis and G. Silva for useful discussions. This work was partially supported by  PICT 2020-03749, PIP 02229, UNLP X910. DHC and VIGR would like to acknowledge support from the ICTP through the Associates Programme. MGF and NAI are supported by fellowships from CONICET (Argentina).  

\appendix

\section{ABJM conventions}
\label{app: ABJM conventions}

The ABJM theory is a three-dimensional ${\cal N}=6$ Chern-Simons-matter theory with gauge group $U(N) \times U(N)$. The field content of the theory is given by gauge fields $\left( A_{\mu} \right)^i_j$ and $( \hat{A}_{\mu} )^{\hat{i}}_{\hat{j}}$ in the adjoint representations of the corresponding gauge groups, bi-fundamental complex scalar fields $( C_I )^{\hat{i}}_j$ and $( \bar{C}^I)^{i}_{\hat{j}}$, and bi-fundamental fermions $( \psi_I )^{i}_{\hat{j}}$ and $( \bar{\psi}^I )^{\hat{i}}_{j}$, where $I=1,\dots,4$.
Supersymmetry transformations of ABJM fields are
\begin{align}
\label{susy ABJ(M) 1}
\delta A_{\mu} &= 2 i \, \bar{\Theta}^{IJ \alpha} \left( \gamma_{\mu} \right)_{\alpha}^{\beta} \left( C_I \psi_{J \beta} + \tfrac{1}{2} \epsilon_{IJKL} \bar{\psi}^K_{\beta} \bar{C}^L \right) \,, \\
\delta \hat{A}_{\mu} &= 2 i \, \bar{\Theta}^{IJ \alpha} \left( \gamma_{\mu} \right)_{\alpha}^{\beta} \left( \psi_{J \beta} C_I + \tfrac{1}{2} \epsilon_{IJKL} \bar{C}^L \bar{\psi}^K_{\beta} \right) \,, \nonumber \\
\delta C_K &= \bar{\Theta}^{IJ \alpha} \epsilon_{IJKL} \bar{\psi}^L_{\alpha} \,, \nonumber \\
\delta \bar{C}^K &= 2 \, \bar{\Theta}^{KL \alpha} \, \psi_{L \alpha} \,, \nonumber \\
\delta \psi^{\beta}_K &= -i \bar{\Theta}^{IJ \alpha} \epsilon_{IJKL} \left( \gamma_{\mu} \right)_{\alpha}^{\beta} D_{\mu} \bar{C}^L + i \,  \bar{\Theta}^{IJ\beta} \epsilon_{IJKL} \left( \bar{C}^L C_P \bar{C}^P - \bar{C}^P C_P  \bar{C}^L \right) + \nonumber \\
& \quad \, + 2i \, \bar{\Theta}^{IJ \beta} \epsilon_{IJML} \bar{C}^M C_K \bar{C}^L \,, \nonumber \\
\label{susy ABJ(M) 6}
\delta \bar{\psi}_{\beta}^K &= -2i \,\bar{\Theta}^{KL \alpha} \left( \gamma_{\mu} \right)_{\alpha \beta} D_{\mu} C_L - 2i \, \bar{\Theta}^{KL}_{\beta} \left( C_L \bar{C}^M C_M - C_M \bar{C}^M  C_L \right) - 4 i \bar{\Theta}^{IJ}_{\beta} C_I \bar{C}^K C_J \,, \nonumber
\end{align}
where the Killing spinors $\bar{\Theta}^{IJ}$ are anti-symmetric in the R-symmetry indices ($\bar{\Theta}^{IJ} =-\bar{\Theta}^{JI}$) and satisfy the reality condition
\begin{equation}
\bar{\Theta}^{IJ}=\left( \Theta_{IJ} \right)^* \,, 
\end{equation}
with
\begin{equation}
\Theta_{IJ} =\frac{1}{2} \epsilon_{IJKL} \bar{\Theta}^{KL} \,.
\end{equation}

\newpage
\section{Supersymmetric Wilson loops in ABJM}
\label{app: review WL}

Let us define the untraced open operator as 
\be
\label{general WL}
 {\cal W}(\tau_1,\tau_2)
 :=
 {\cal P}\exp\left( {i \int_{\tau_1}^{\tau_2}
 {\cal L}(\tau) d\tau}\right)\,,
\ee
In order to construct BPS Wilson loops, one has to demand that, under a supersymmetry transformation \cite{Drukker:2009hy,Lee:2010hk},
\be
\delta {\cal L}= {\cal D}_{\tau} \Lambda := \partial_{\tau} \Lambda +i \{ {\cal L}, \Lambda ] \,,
\label{susycondition}
\ee
where $\Lambda$ is some supermatrix. Thus, under a finite supersymmetry transformation 
\be
\label{WL finite transformation}
{\cal W} (\tau_1,\tau_2)  \to \exp[-i \Lambda (\tau_1)] (\tau_1) {\cal W} (\tau_1,\tau_2) 
\exp[i \Lambda (\tau_2)] \,,
\ee

For the 1/2 BPS Wilson line \eqref{WLdef}-\eqref{Ldef}, the $\Lambda$ supermatrix is
\be
\label{lambda straight WL}
\Lambda = \left(
\begin{array}{cc}
0 & g_1
\\
\bar g_2 & 0
\end{array}\right)\,,
\ee
with\cite{Cardinali:2012ru}
\begin{align} 
\label{g1}
g_1 & = 2 \, \eta \, \bar{\Theta}^{12}_{+} C_2 
+\eta \, \bar{\Theta}^{13}_{+} C_3 +\eta \, \bar{\Theta}^{14}_{+} C_4 \,,
\\
\bar g_2 & =  
\bar{\eta} \, \bar{\Theta}^{23}_{-} \bar{C}^4
-\bar{\eta} \, \bar{\Theta}^{24}_{-} \bar{C}^3 
+\bar{\eta} \, \bar{\Theta}^{34}_{-} \bar{C}^2\,.
\end{align}
The Wilson loop defined by the parametrization \eqref{WLdef}-\eqref{Ldef} is invariant under an $SU(1,1|3)\subset OSp(6|4)$. 
All possible insertions along the 1/2 BPS Wilson line should be organized  into representations of $SU(1,1|3)$. A detailed discussion on the corresponding representations can be found in \cite{Bianchi:2017ozk,Bianchi:2020hsz,Agmon:2020pde}. These representations can be short or long, depending on
whether the superprimary is annihilated or not by some of the supersymmetries. The analysis of the supersymmetry of Wilson loops in the ABJM model is a bit more subtle, as  the untraced loop is covariant rather than invariant. To determine the supersymmetry preserved by an insertion ${\cal O}$ along the Wilson line, it is then convenient to introduce  a \textit{covariant} supersymmetric transformation  \cite{Gorini:2022jws} as
\begin{equation}
\label{covariant susy transformation}
\delta^{\rm cov} {\cal O} :=\delta {\cal O} - i \{ {\cal O}, \Lambda] \,.
\end{equation}

For example, consider the off-diagonal insertions of the form
\be 
{\cal O}_I = 
\left(
\begin{array}{cc}
 0 &  C_I 
\\
0 &  0
\end{array}\right)\,.
\label{susyinsertion00a}
\ee
In the case of ${\cal O}_1$, it is straightforward to check that $\delta^{\rm cov} {\cal O}_1 = 0$ when the supersymmetry transformation is generated by any of the parameters $\bar{\Theta}^{1J}_{+}$ appearing in \eqref{g1},
which indicates that ${\cal O}_1$ is the primary of 1/2 BPS short multiplet. Its descendants and those of its anti-chiral conjugate
$\bar {\cal O}^1$ include six exactly marginal insertions. When considered as deformations of the dCFT$_1$, they trigger unitary rotations of the coset $SU(4)/(SU(3)\times U(1))$. In the dual picture, they correspond to rotations of the point where the string is fixed at the $\mathbb{CP}^3$.

Interestingly, the remaining ${\cal O}_I$ insertions with $I=2,3,4$ are not primaries but descendants of the constant operator insertion \cite{Gorini:2022jws}
\begin{equation}
    {\cal T} = \left(
\begin{array}{cc}
\mathds{1} & 0
\\
0 &  -\mathds{1}
\end{array}\right)\,.
\end{equation}

For example, ${\cal O}_2$ is obtained when we transform ${\cal T}$ with the supersymmetry generated by $\bar\Theta^{12}_+$. It is interesting to go down another step in the multiplet. Schematically,
\begin{equation}
    {\cal T} \quad \overset{\bar\Theta^{12}_+} {\longrightarrow}\quad {O}_2 \quad \overset{\bar{\Theta}^{34}_{-}} {\longrightarrow}\quad \left(
\begin{array}{cc}
2C_2 \bar C^2& {\eta}\bar\psi^1_+
\\
0 &  \bar 2 C^2 C_2
\end{array}\right)\,,
\end{equation}
or by going down differently, 
\begin{equation}
    {\cal T} \quad \overset{\bar{\Theta}^{34}_{-}} {\longrightarrow}\quad {\cal \bar O}^2 \quad \overset{\bar\Theta^{12}_+} {\longrightarrow}\quad \left(
\begin{array}{cc}
2C_2 \bar C^2 & 0
\\
\psi_1^+ \bar{\eta}&  \bar 2 C^2 C_2
\end{array}\right)\,,
\end{equation}

Thus, a combination of descendants of the constant insertion ${\cal T}$ gives rise to
\be
{\cal O} =  \begin{pmatrix}
 4 C_2 \bar{C}^2 &  \eta \bar{\psi}^{1}_{+}
\\
  \psi^{+}_{1} \bar{\eta} & 4 \bar{C}^2 C_2
\end{pmatrix}
\,,
\label{insertion}
\ee
which is the operator triggering one of the possible supersymmetric deformations of the 1/2 BPS line defect CFT \cite{Ouyang:2015iza,Ouyang:2015bmy}. The scaling dimension $\Delta_{\cal O}$  would determine the nature of the deformation of the line defect.

\section{Scattering matrix}
\label{convencionessmatrix}
We will discuss the conventions of the scattering matrix between two fundamental magnons. The S-matrix is a map between {\sl in} and {\sl out} asymptotic states. Given the tensor product of two fundamental representations of $SU(2|2)$ ${\cal V}_1\otimes{\cal V}_1$, each of which is specified by the quantum numbers $(\zeta_1,x^\pm)$ and $(\zeta_2,y^\pm)$, we say it represents an {\sl in} state when $\zeta_2=\zeta_1\tfrac{x^+}{x^-}$ and it represents an {\sl out} state when $\zeta_1=\zeta_2\tfrac{y^+}{y^-}$.  The action of the S-matrix is
such that $|{\rm out}\rangle = {\mathbf S} |{\rm in}\rangle $. More concretely,
\begin{equation}
{\mathbf S}	\,|\chi_I(x^\pm)\chi_J(y^\pm)\rangle_{\rm in} 
=  {S}_{IJ}^{KL}(x^\pm,y^\pm)|\chi_K(x^\pm)\chi_L(y^\pm)\rangle_{\rm out}\,.
\end{equation}

The S-matrix components can be written as ${S}_{IJ}^{KL} = S_0\, \hat {S}_{IJ}^{KL}$ and fixed up to an overall scalar factor by demanding that it commutes with the $SU(2|2)$ generators, namely 
 \begin{alignat}{4}
    \hat S_{ab}^{cd} 
        &= \frac{1}{2}(A-B)\delta_a^c\delta_b^d + \frac{1}{2}(A+B)\delta_a^d\delta_b^c,
        &\qquad
    \hat S_{ab}^{\alpha\beta} 
        &= \frac{1}{2}\epsilon_{ab}\epsilon^{\alpha\beta}C, \\
   \hat S_{\alpha\beta}^{\rho\sigma} 
        &= \delta_{\alpha}^{\rho}\delta_{\beta}^{\sigma}\frac{D-E}{2} + \delta_{\alpha}^{\sigma}\delta_{\beta}^{\rho}\frac{D+E}{2},
        &\qquad
    \hat S_{a\alpha}^{b\beta} 
        &= \delta_a^b\delta_{\alpha}^{\beta}G, \\
    \hat S_{a\alpha}^{\beta b} 
        &= \delta_a^b\delta_{\alpha}^{\beta}H,
        &\qquad
    \hat S_{\alpha a}^{b\beta} 
        &= \delta_a^b\delta_{\alpha}^{\beta}K, \\
    \hat S_{\alpha a}^{\beta b} 
        &= \delta_a^b\delta_{\alpha}^{\beta}L,
        &\qquad
    \hat S_{\alpha\beta}^{ab} 
        &= \frac{1}{2}\epsilon_{\alpha\beta}\epsilon^{ab}F
\end{alignat}
where
\begin{equation}
\begin{aligned}
A &= \frac{x^- - y^+}{x^+ - y^-} \,
\frac{\eta_1 \; \eta_2}
{\tilde{\eta}_2 \;\tilde{\eta}_1}, \\
B &= \frac{y^+ - x^-}{y^- - x^+}
\bigg(1 - 2 \,
\frac{1 - \frac{1}{y^- x^+}}{1 - \frac{1}{y^+ x^+}} \,
\frac{y^- - x^-}{y^+ - x^-}\bigg)
\frac{\eta_1 \; \eta_2}
{\tilde{\eta}_2 \;\tilde{\eta}_1}, \\
C &= \frac{2i}{x^+ y^+} \,
\frac{y^- - x^-}{y^- - x^+} \,
\frac{1}{1 - \frac{1}{y^+ x^+}} \,
\frac{\eta_1 \; \eta_2}{\zeta}, \\
D &= -1, \\
E &= -\!\bigg(1 - 2 \,
\frac{1 - \frac{1}{y^+ x^-}}{1 - \frac{1}{y^- x^-}} \,
\frac{y^+ - x^+}{y^- - x^+} \bigg), \\
F &= \frac{2i (x^+ - x^-)(y^+ - y^-)}{x^- y^-} \,
\frac{y^+ - x^+}{y^- - x^+} \,
\frac{1}{1 - \frac{1}{y^- x^-}} \,
\frac{\zeta}{\tilde{\eta}_2 \;\tilde{\eta}_1}, \\
G &= \frac{x^+ - y^+}{x^+ - y^-} \,
\frac{\eta_1}{\tilde{\eta}_1}, \\
H &= \frac{y^- - y^+}{x^+ - y^-} \,
\frac{\eta_1}{\tilde{\eta}_2}, \\
K &= \frac{x^- - x^+}{x^+ - y^-} \,
\frac{\eta_2}
{\tilde{\eta}_1}, \\
L &= \frac{x^- - y^-}{x^+ - y^-} \,
\frac{\eta_2}
{\tilde{\eta}_2}.
\end{aligned}
\end{equation}
and
\begin{equation}
    \eta_1=\eta(\zeta,x^{\pm}),\quad\tilde{\eta}_1=\eta(\zeta \tfrac{y^+}{y^-},x^{\pm}),\quad\eta_2=\eta(\zeta \tfrac{x^+}{x^-},y^{\pm}),\quad\tilde{\eta}_2=\eta(\zeta,y^{\pm})
\end{equation}
In the ABJM spin chain, the bulk dressing factor between magnons of the same type is
\begin{equation}
\label{S0AA}
S_0(x^\pm,y^\pm) = 
\frac{x^+}{x^-} \,
\frac{1-\frac{1}{x^- y^+}}{1-\frac{1}{x^+ y^-}}\, \frac1{\sigma(x^\pm,y^\pm)} \,,
\end{equation}
and $\sigma(x^\pm,y^\pm)$ is the BES dressing factor \cite{Beisert:2006ez}.
The bulk dressing factor between magnons of different types is
\begin{equation}
\label{S0AB}
\tilde{S}_0(x^\pm,y^\pm) = \frac
{x^+}{x^-}\,  \frac{x^+-y^-}{x^--y^+} \,
\frac1{\sigma(x^\pm,y^\pm)} \,.
\end{equation}

\section{Functions in the reflection matrix solution}
\label{lista}
In order to have expressions as compact as possible, we fix $G_B^4 = 1/{\eta_2}$,
\begin{equation}
\begin{aligned}
A_B = & \  \frac{x^+ - x^-(1 + x^{+2}) + 2 i x^{-3}x^+\kappa + x^{-2}\bigl(x^+ -i \kappa + i x^-(x^+ + i\kappa)\kappa\bigr)}
{  x^- (x^- + x^+) ( \kappa+\tfrac{1}{g} ) }
\frac{\sqrt{f}}{\eta}\,,
\\
B_B = &\  A_B +\frac{2 (x^{-} - x^{+})(1 + x^{+2})\kappa}
{x^+(x^+-x^- +i \kappa)}
\frac{\sqrt{f}}{\eta} \,,
\qquad
C_B^1 = \frac{2\kappa \bigl(x^- - x^+(1 - i x^+\kappa)\bigr)}
{x^+ (x^+-x^- +i \kappa)} \eta_2 
\,,
\\
C_B^2 = &- 
\frac{i\kappa (x^+ + x^{-2} x^+ + x^{+3} - x^- 
(x^+(i\kappa + 2 x^+) + 1))}
{x^{+^2}(x^+-x^- + i\kappa)(\kappa+\tfrac{1}{g})} \Sigma\, \eta_2\,,\\
D_B^1 = & \sqrt{\frac{x^-}{x^+}}
\frac{x^+ + x^- 
\left( -1 + (x^- - x^+) x^+ \right) 
-i x^{+2}
\left( 1 + x^{-2} - 2 x^- x^+ \right) \kappa - x^- x^{+2} \kappa^2}{ x^+ (x^- + x^+) (\kappa+\tfrac{1}{g} ) } \frac{\sqrt{f}}{\eta}\,,
\\
D_B^2 = &\ -
\sqrt{\frac{x^-}{x^+}}\frac{i\left(x^{-} - x^{+}\right) ( (x^- + x^+)\kappa+\tfrac{1}{g} x^- )}
{2 x^- (x^+ - x^- +i \kappa)
  (\kappa+\tfrac{1}{g}) }
 \frac{\sqrt{f}\,\Sigma}{\eta}
\,,
 \\
D_B^3 = &\sqrt{\frac{x^+}{x^-}}\Bigl(A_B-\frac{
\bigl((x^- - x^+)^2 -x^{+2} \kappa^2\bigr)
\bigl(x^+ - x^- (1 + x^+ (x^+ - x^- +i\kappa))\bigr)}{x^- x^{+2} (x^+ -x^- +i\kappa) (\kappa+\frac{1}{g}) }\frac{\sqrt{f}}{\eta}\Bigr),\\
D_B^5 = &-\sqrt{\frac{x^-}{x^+}}\Bigl( \frac{i \;x^- \,\kappa\,\bigl((\tfrac{1}{x^+} - \tfrac{1}{x^-})^2\bigl(1 + x^+(x^+ + x^-(-2 + x^- x^+))\bigr) - \frac{2\kappa}{g} -\kappa^2\bigr)}
{(x^+ - x^- +i \kappa)(\kappa +\frac{1}{g} )}\frac{\sqrt{f}}{ \eta}+A_B \Bigr)\;,
\\
H_B^3 = &\frac{x^-}{x^+}\bigl(A_B-i\frac{
    x^+( x^{-} -x^{+} ) \kappa
}{(\kappa+\tfrac{1}{g}) }  
\frac{\sqrt{f}}{\eta}
\bigr)\,,
\qquad
L_B^4 =- \sqrt{\frac{x^+}{x^-}}\frac{i\,(x^- - x^+)\,\kappa\,\eta_2 f}{\eta^2}\,,
\\
H_B^4 = &
\frac{x^{-2} +i \tfrac{x^{+2}}{x^-}\kappa - \tfrac{x^{-}}{x^+}\bigl(1 + x^{+2} +i x^+\kappa -2i x^{+3}\kappa\bigr) + \bigl(1 - ix^+\kappa(1 + x^{+2} +i x^+\kappa)\bigr)}{ (x^- + x^+) (\kappa+\frac{1}{g} ) }\frac{\sqrt{f}}{\eta}
\,,\\ 
K_B^3 = & 
\frac{\sqrt{\frac{x^-}{x^+}}\bigl(i x^{-3}x^+\kappa + x^+(1 -i x^+\kappa) + x^{-2}x^+(1 -i x^+\kappa) - x^-(1 + x^{+2} +i x^+\kappa)(1 -i x^+\kappa)\bigr)}{ x^- (x^- + x^+) (\kappa+\frac{1}{g})} \frac{\sqrt{f}}{\eta}\;,
\\
 K_B^{4} = & 
\frac{\sqrt{\frac{x^+}{x^-}}\Bigl( i x^{-3}\kappa - x^{-}(x^+ -i \kappa) + \bigl(1 + x^{+2} +i x^+\kappa\bigr) +\tfrac{x^{+}}{x^-}-i \tfrac{x^{-2}}{x^+}\kappa\bigl(1 + x^+(2x^+ +i \kappa)\bigr)\Bigr)}
{(x^- + x^+) (\kappa+\frac{1}{g} ) }\frac{\sqrt{f}}{\eta}
\,,\\
    \end{aligned}
    \nonumber
\end{equation}
where
\begin{equation}
 \Sigma =-\kappa^2 -\frac{1}{g}\kappa   
+\frac{i}{g}
\sqrt{\kappa[-\kappa+ 2 g(-\kappa^2+2)+\kappa g^2(-\kappa^2+4)]}\,.
\end{equation}
The remaining functions can be written using the relations:
\begin{alignat}{6}
F_B^1 = & \  \frac{1}{\eta_2^2 \kappa} C_B^1\,,
 \qquad &
 F_B^2 = &\ - \frac{4(\kappa+\tfrac{1}{g})}{\eta_2^2 \Sigma^2} C_B^2\,, 
 \qquad &
 G_B^3 = &\ -\frac{2i\,x^{-}\kappa}
{\Sigma}  G_B^4\,,
 \\
 D_B^6 = & \ -\frac{x^+}{x^-}  D_B^1\,,
 \qquad &\
 D_B^4 =  & \frac{4\kappa(\kappa+\tfrac{1}{g})}{\Sigma^2} D_B^2\,,
 \qquad &
 L_B^3 = &\ - \frac{i x^- \, \Sigma}{\kappa+\tfrac{1}{g}}L_B^4\,.
\end{alignat}


\begin{thebibliography}{99}

\bibitem{Drukker:2006xg}
N.~Drukker and S.~Kawamoto,
``Small deformations of supersymmetric Wilson loops and open spin-chains,''
JHEP \textbf{07} (2006), 024
\href{https://arxiv.org/abs/hep-th/0604124}{\tt arXiv:hep-th/0604124}.



\bibitem{Drukker:2012de}
N.~Drukker,
`Integrable Wilson loops,''
JHEP \textbf{10} (2013), 135
\href{https://arxiv.org/abs/1203.1617}{\tt arXiv:1203.1617 [hep-th]}.

\bibitem{Correa:2012hh}
D.~Correa, J.~Maldacena and A.~Sever,
``The quark anti-quark potential and the cusp anomalous dimension from a TBA equation,''
JHEP \textbf{08} (2012), 134
\href{https://arxiv.org/abs/1203.1913}{\tt arXiv:1203.1913 [hep-th]}.

\bibitem{Gromov:2013pga}
N.~Gromov, V.~Kazakov, S.~Leurent and D.~Volin,
``Quantum Spectral Curve for Planar $\mathcal{N} = 4$ Super-Yang-Mills Theory,''
Phys. Rev. Lett. \textbf{112} (2014) no.1, 011602
\href{https://arxiv.org/abs/1305.1939}{\tt arXiv:1305.1939 [hep-th]}.



\bibitem{Gromov:2015dfa}
N.~Gromov and F.~Levkovich-Maslyuk,
``Quantum Spectral Curve for a cusped Wilson line in $ \mathcal{N}=4 $ SYM,''
JHEP \textbf{04} (2016), 134
\href{https://arxiv.org/abs/1510.02098}{\tt arXiv:1510.02098 [hep-th]}.



\bibitem{Liendo:2018ukf}
P.~Liendo, C.~Meneghelli and V.~Mitev,
``Bootstrapping the half-BPS line defect,''
JHEP \textbf{10} (2018), 077
\href{https://arxiv.org/abs/1806.01862}{\tt arXiv:1806.01862 [hep-th]}.



\bibitem{Cavaglia:2021bnz}
A.~Cavagli{\`a}, N.~Gromov, J.~Julius and M.~Preti,
``Integrability and conformal bootstrap: One dimensional defect conformal field theory,''
Phys. Rev. D \textbf{105} (2022) no.2, L021902
\href{https://arxiv.org/abs/2107.08510}{\tt arXiv:2107.08510 [hep-th]}.



\bibitem{Correa:2023lsm}
D.~H.~Correa, V.~I.~Giraldo-Rivera and M.~Lagares,
``Integrable Wilson loops in ABJM: a Y-system computation of the cusp anomalous dimension,''
JHEP \textbf{06} (2023), 179
\href{https://arxiv.org/abs/2304.01924}{\tt arXiv:2304.01924 [hep-th]}.



\bibitem{Beisert:2006fmy}
N.~Beisert,
``The S-matrix of AdS / CFT and Yangian symmetry,''
PoS \textbf{SOLVAY} (2006), 002
\href{https://arxiv.org/abs/0704.0400}{\tt arXiv:0704.0400 [nlin.SI]}.


\bibitem{Arutyunov:2009mi}
G.~Arutyunov, M.~de Leeuw and A.~Torrielli,
``The Bound State S-Matrix for AdS(5) x S**5 Superstring,''
Nucl. Phys. B \textbf{819} (2009), 319-350
\href{https://arxiv.org/abs/0902.0183}{\tt arXiv:0902.0183 [hep-th]}.


\bibitem{Loebbert:2016cdm}
F.~Loebbert,
``Lectures on Yangian Symmetry,''
J. Phys. A \textbf{49} (2016) no.32, 323002
\href{https://arxiv.org/abs/1606.02947 }{\tt arXiv:1606.02947 [hep-th]}.


\bibitem{Loebbert:2024qbw}
F.~Loebbert and H.~Mathur,
``The Feyn-structure of Yangian symmetry'',JHEP \textbf{01} (2025), 112
\href{https://arxiv.org/abs/2410.11936}{\tt arXiv:2410.11936 [hep-th]}.



\bibitem{Loebbert:2025abz}
F.~Loebbert, L.~R{\"u}enaufer and S.~F.~Stawinski,
``Nonlocal Symmetries of Planar Feynman Integrals,''
Phys. Rev. Lett. \textbf{135} (2025) no.15, 151603
\href{https://arxiv.org/abs/2505.05550}{\tt arXiv:2505.05550 [hep-th]}.



   
\bibitem{MacKay:2001bh}
N.~J.~MacKay and B.~J.~Short,
``Boundary scattering, symmetric spaces and the principal chiral model on the half line,''
Commun. Math. Phys. \textbf{233} (2003), 313-354 [erratum: Commun. Math. Phys. \textbf{245} (2004), 425-428]
\href{https://arxiv.org/abs/hep-th/0104212}{arXiv:hep-th/0104212 [hep-th]}.

\bibitem{Delius:2001he}
G.~W.~Delius, N.~J.~MacKay and B.~J.~Short,
``Boundary remnant of Yangian symmetry and the structure of rational reflection matrices,''
Phys. Lett. B \textbf{522} (2001), 335-344
[erratum: Phys. Lett. B \textbf{524} (2002), 401-401]
\href{https://arxiv.org/abs/hep-th/0109115}{arXiv:hep-th/0109115 [hep-th]}.




\bibitem{Hofman:2007xp}
D.~M.~Hofman and J.~M.~Maldacena,
``Reflecting magnons,''
JHEP \textbf{11} (2007), 063
\href{https://arxiv.org/abs/0708.2272}{\tt arXiv:0708.2272 [hep-th]}.

\bibitem{Ahn:2010xa}
C.~Ahn and R.~I.~Nepomechie,
``Yangian symmetry and bound states in AdS/CFT boundary scattering,''
JHEP \textbf{05} (2010), 016
\href{https://arxiv.org/abs/1003.3361}{\tt arXiv:1003.3361 [hep-th]}.


\bibitem{MacKay:2010zb}
N.~MacKay and V.~Regelskis,
``On the reflection of magnon bound states,''
JHEP \textbf{08} (2010), 055
\href{https://arxiv.org/abs/1006.4102}{\tt arXiv:1006.4102 [hep-th]}.



\bibitem{MacKay:2010ey}
N.~MacKay and V.~Regelskis,
``Yangian symmetry of the Y=0 maximal giant graviton,''
JHEP \textbf{12} (2010), 076
\href{https://arxiv.org/abs/1010.3761}{\tt arXiv:1010.3761 [hep-th]}.


\bibitem{Palla:2011eu}
L.~Palla,
``Yangian symmetry of boundary scattering in AdS/CFT and the explicit form of bound state reflection matrices,''
JHEP \textbf{03} (2011), 110
\href{https://arxiv.org/abs/1102.0122}{\tt arXiv:1102.0122 [hep-th]}.


\bibitem{Correa:2008av}
D.~H.~Correa and C.~A.~S.~Young,
``Reflecting magnons from D7 and D5 branes,''
J. Phys. A \textbf{41} (2008), 455401
\href{https://arxiv.org/abs/0808.0452}{\tt arXiv:0808.0452 [hep-th]}.


\bibitem{Chen:2018sbp}
H.~H.~Chen, H.~Ouyang and J.~B.~Wu,
``Open Spin Chains from Determinant Like Operators in ABJM Theory,''
Phys. Rev. D \textbf{98} (2018) no.10, 106012
\href{https://arxiv.org/abs/1809.09941}{\tt arXiv:1809.09941 [hep-th]}.


\bibitem{Aharony:2008ug}
O.~Aharony, O.~Bergman, D.~L.~Jafferis and J.~Maldacena,
``N=6 superconformal Chern-Simons-matter theories, M2-branes and their gravity duals,''
JHEP \textbf{10} (2008), 091
\href{https://arxiv.org/abs/0806.1218}{\tt arXiv:0806.1218 [hep-th]}.



\bibitem{Drukker:2009hy}
N.~Drukker and D.~Trancanelli,
``A Supermatrix model for N=6 super Chern-Simons-matter theory,''
JHEP \textbf{02} (2010), 058,
\href{https://arxiv.org/abs/0912.3006}{\tt arXiv:0912.3006 [hep-th]}.


\bibitem{Lee:2010hk}
K.~M.~Lee and S.~Lee,
``1/2-BPS Wilson Loops and Vortices in ABJM Model,'' JHEP \textbf{09} (2010), 004
\href{https://arxiv.org/abs/1006.5589}{\tt arXiv:1006.5589 [hep-th]}.


\bibitem{Minahan:2008hf}
J.~A.~Minahan and K.~Zarembo,
``The Bethe ansatz for superconformal Chern-Simons,''
JHEP \textbf{09} (2008), 040
\href{https://arxiv.org/abs/0806.3951}{\tt arXiv:0806.3951 [hep-th]}.

\bibitem{Gaiotto:2008cg}
D.~Gaiotto, S.~Giombi and X.~Yin,
``Spin Chains in N=6 Superconformal Chern-Simons-Matter Theory,''
JHEP \textbf{04} (2009), 066,
\href{https://arxiv.org/abs/0806.4589}{\tt arXiv:0806.4589 [hep-th]}.

\bibitem{Gromov:2008qe}
N.~Gromov and P.~Vieira,
``The all loop AdS4/CFT3 Bethe ansatz,''
JHEP \textbf{01} (2009), 016
\href{https://arxiv.org/abs/0807.0777}{\tt arXiv:0807.0777 [hep-th]}.

\bibitem{Gromov:2014eha}
N.~Gromov and G.~Sizov,
``Exact Slope and Interpolating Functions in N=6 Supersymmetric Chern-Simons Theory,''
Phys. Rev. Lett. \textbf{113} (2014) no.12, 121601
\href{https://arxiv.org/abs/1403.1894}{\tt arXiv:1403.1894 [hep-th]}.


\bibitem{Beisert:2005tm}
N.~Beisert,
``The SU(2|2) dynamic S-matrix,''
Adv. Theor. Math. Phys. \textbf{12} (2008), 945-979,
\href{https://arxiv.org/abs/hep-th/0511082}{arXiv:hep-th/0511082 [hep-th]}.

\bibitem{Beisert:2006ez}
N.~Beisert, B.~Eden and M.~Staudacher,
``Transcendentality and Crossing,''
J. Stat. Mech. \textbf{0701} (2007), P01021,
\href{https://arxiv.org/abs/hep-th/0610251}{\tt arXiv:hep-th/0610251 [hep-th]}.


\bibitem{Ahn:2008aa}
C.~Ahn and R.~I.~Nepomechie,
``N=6 super Chern-Simons theory S-matrix and all-loop Bethe ansatz equations,''
JHEP \textbf{09} (2008), 010,
\href{https://arxiv.org/abs/0807.1924}{\tt arXiv:0807.1924 [hep-th]}.


\bibitem{Drukker:2019bev}
N.~Drukker, D.~Trancanelli, L.~Bianchi, M.~S.~Bianchi, D.~H.~Correa, V.~Forini, L.~Griguolo, M.~Leoni, F.~Levkovich-Maslyuk and G.~Nagaoka, \textit{et al.}
``Roadmap on Wilson loops in 3d Chern{\textendash}Simons-matter theories,''
J. Phys. A \textbf{53} (2020) no.17, 173001
\href{https://arxiv.org/abs/1910.00588}{\tt arXiv:1910.00588 [hep-th]}.

\bibitem{Bombardelli:2009xz}
D.~Bombardelli, D.~Fioravanti and R.~Tateo,
``TBA and Y-system for planar AdS(4)/CFT(3),''
Nucl. Phys. B \textbf{834} (2010), 543-561
\href{https://arxiv.org/abs/0912.4715}{\tt arXiv:0912.4715 [hep-th]}.


\bibitem{Correa:2009mz}
D.~H.~Correa and C.~A.~S.~Young,
``Finite size corrections for open strings/open chains in planar AdS/CFT,''
JHEP \textbf{08} (2009), 097
\href{https://arxiv.org/abs/0905.1700}{\tt arXiv:0905.1700 [hep-th]}.

\bibitem{Drinfeld:1985rx}
V.~G.~Drinfeld,
``Hopf algebras and the quantum Yang-Baxter equation,''
Sov. Math. Dokl. \textbf{32} (1985), 254-258.

\bibitem{Correa:2009dm}
D.~H.~Correa and C.~A.~S.~Young,
``Asymptotic Bethe equations for open boundaries in planar AdS/CFT,''
J. Phys. A \textbf{43} (2010), 145401
\href{https://arxiv.org/abs/0912.0627}{\tt arXiv:0912.0627 [hep-th]}.


\bibitem{Ghoshal:1993tm}
S.~Ghoshal and A.~B.~Zamolodchikov,
``Boundary S matrix and boundary state in two-dimensional integrable quantum field theory,''
Int. J. Mod. Phys. A \textbf{9} (1994), 3841-3886
[erratum: Int. J. Mod. Phys. A \textbf{9} (1994), 4353]
\href{https://arxiv.org/abs/hep-th/9306002}{\tt arXiv:hep-th/9306002 [hep-th]}.

\bibitem{Bajnok:2004jd}
Z.~Bajnok and A.~George,
``From defects to boundaries,''
Int. J. Mod. Phys. A \textbf{21} (2006), 1063-1078
\href{https://arxiv.org/abs/hep-th/0404199}{\tt arXiv:hep-th/0404199 [hep-th]}.


\bibitem{Gorini:2022jws}
N.~Gorini, L.~Griguolo, L.~Guerrini, S.~Penati, D.~Seminara and P.~Soresina,
``Constant primary operators and where to find them: the strange case of BPS defects in ABJ(M) theory,''
JHEP \textbf{02} (2023), 013,
\href{https://arxiv.org/abs/2209.11269}{\tt arXiv:2209.11269 [hep-th]}.


\bibitem{Kim:2017sju}
M.~Kim, N.~Kiryu, S.~Komatsu and T.~Nishimura,
``Structure Constants of Defect Changing Operators on the 1/2 BPS Wilson Loop,''
JHEP \textbf{12} (2017), 055
\href{https://arxiv.org/abs/2209.11269}{\tt arXiv:1710.07325 [hep-th]}.

\bibitem{Griguolo:2012iq}
L.~Griguolo, D.~Marmiroli, G.~Martelloni and D.~Seminara,
``The generalized cusp in ABJ(M) N = 6 Super Chern-Simons theories,''
JHEP \textbf{05} (2013), 113
\href{https://arxiv.org/abs/1208.5766}{\tt arXiv:1208.5766 [hep-th]}.


\bibitem{Ouyang:2015iza}
H.~Ouyang, J.~B.~Wu and J.~j.~Zhang,
``Novel BPS Wilson loops in three-dimensional quiver Chern{\textendash}Simons-matter theories,''
Phys. Lett. B \textbf{753} (2016), 215-220
\href{https://arxiv.org/abs/1510.05475}{\tt arXiv:1510.05475 [hep-th]}.

\bibitem{Ouyang:2015bmy}
H.~Ouyang, J.~B.~Wu and J.~j.~Zhang,
``Construction and classification of novel BPS Wilson loops in quiver Chern{\textendash}Simons-matter theories,''
Nucl. Phys. B \textbf{910} (2016), 496-527
\href{https://arxiv.org/abs/1511.02967}{\tt arXiv:1511.02967 [hep-th]}.


\bibitem{Correa:2019rdk}
D.~H.~Correa, V.~I.~Giraldo-Rivera and G.~A.~Silva,
``Supersymmetric mixed boundary conditions in AdS$_{2}$ and DCFT$_{1}$ marginal deformations,''
JHEP \textbf{03} (2020), 010
\href{https://arxiv.org/abs/1910.04225}{\tt arXiv:1910.04225 [hep-th]}.

\bibitem{Castiglioni:2022yes}
L.~Castiglioni, S.~Penati, M.~Tenser and D.~Trancanelli,
``Interpolating Wilson loops and enriched RG flows,''
JHEP \textbf{08} (2023), 106
\href{https://arxiv.org/abs/2211.16501}{\tt arXiv:2211.16501 [hep-th]}


\bibitem{Castiglioni:2023uus}
L.~Castiglioni, S.~Penati, M.~Tenser and D.~Trancanelli,
``Wilson loops and defect RG flows in ABJM,''
JHEP \textbf{06} (2023), 157
\href{https://arxiv.org/abs/2305.01647}{\tt arXiv:2305.01647 [hep-th]}


\bibitem{Cardinali:2012ru}
V.~Cardinali, L.~Griguolo, G.~Martelloni and D.~Seminara,
``New supersymmetric Wilson loops in ABJ(M) theories,''
Phys. Lett. B \textbf{718} (2012), 615-619
\href{https://arxiv.org/abs/1209.4032}{\tt arXiv:1209.4032 [hep-th]}.


\bibitem{Bianchi:2017ozk}
L.~Bianchi, L.~Griguolo, M.~Preti and D.~Seminara,
``Wilson lines as superconformal defects in ABJM theory: a formula for the emitted radiation,''
JHEP \textbf{10} (2017), 050,
\href{https://arxiv.org/abs/1706.06590}{\tt arXiv:1706.06590 [hep-th]}.


\bibitem{Agmon:2020pde}
N.~B.~Agmon and Y.~Wang,
``Classifying Superconformal Defects in Diverse Dimensions Part I: Superconformal Lines,''
\href{https://arxiv.org/abs/2009.06650}{\tt arXiv:2009.06650 [hep-th]}.


\bibitem{Bianchi:2020hsz}
L.~Bianchi, G.~Bliard, V.~Forini, L.~Griguolo and D.~Seminara,
``Analytic bootstrap and Witten diagrams for the ABJM Wilson line as defect CFT$_{1}$,''
JHEP \textbf{08} (2020), 143,
\href{https://arxiv.org/abs/2004.07849}{\tt arXiv:2004.07849 [hep-th]}.

\end{thebibliography}
\end{document}